# Nonlinear Terahertz Resonances from Ballistic Electron Funnelling

Hue T.B. Do[1,2,4], Gregory K. Ngirmang[3], Wu Lin[3]*, Michel Bosman[1,4]*

[1]Department of Materials Science and Engineering, National University of Singapore, 9 Engineering Drive 1, 117575, Singapore.

[2]NUS Graduate School - Integrative Sciences and Engineering Programme (ISEP), National University of Singapore, 21 Lower Kent Ridge Road, 119077, Singapore.

[3]Department of Science, Mathematics and Technology, Singapore University of Technology and Design, 8 Somapah Road, 487372, Singapore.

[4]Institute of Materials Research and Engineering, Agency for Science, Technology and Research (A*STAR), 2 Fusionopolis Way, 138634, Singapore.

*lin_wu@sutd.edu.sg; *msemb@nus.edu.sg

**Abstract**

We introduce a new mechanism for second-harmonic generation through geometrically rectifying—funnelling—ballistic electrons in THz optical resonators. Our resonant rectifiers inherently act as second-order harmonic generators, rectifying currents without the presence of a potential barrier. Particle-in-cell simulations reveal that femtosecond electron-surface scattering plays a critical role in this process. We differentiate electron funnelling from nonlocal plasmonic drag and bulk Dirac anharmonicity, showing that funnelling can reduce the required field intensity for second-harmonic generation by 3-4 orders of magnitude. We provide design guidelines for generating funnelling-induced second-harmonic generation, including resonance mode matching and materials selection. This approach offers a practical pathway for low-field, geometrically tunable THz upconversion and rectification, operating from sub-10 THz to multiple tens of THz in graphene.

# Introduction

Ballistic rectifiers are devices in which ballistic transport of charge carriers achieves current rectification. This can be achieved when the inelastic mean-free-path of the charge carriers is larger than their geometrical confinement. In that case, they bounce off the surface like billiard balls, and by cleverly designing their confined geometries, current rectification can be achieved[1–4]. Such a rectification mechanism is attractive as it does not require any doped junction or potential barrier, allowing the device to operate without a threshold voltage. Rectification is achieved purely by engineering the geometry of nanopatterned materials.

Resonant rectifiers inherently act as second-order harmonic generators. However, the potential for geometrically controlling high harmonic generation with ballistic electrons has not been explored in previous work on ballistic rectifiers. This presents an interesting opportunity. By designing simple rectifying geometries as optical resonators, ballistic electrons can couple with light, setting up plasmonic harmonic resonances. This not only enables light-induced current rectification but also allows for frequency doubling when energy transfer occurs between low- and high-frequency plasmon modes. Such optical rectifiers and frequency doublers could open up new applications in THz technology, including mid-IR photodiodes and infrared radiation harvesting for power generation.

Nonlinear plasmonics with ballistic electrons have been studied in 2D electron gases, particularly in short-channel field-effect transistors exhibiting Dyakonov-Shur (DS) instabilities, where plasmons in a Fabry-Perot cavity are rectified due to boundary asymmetries between the source and drain[5–7]. In graphene nanostructures, nonlinear plasmonics arise from the anharmonicity of Dirac electrons[8,9], enhanced by large local plasmonic fields[10,11]. However, previous studies have overlooked electron-surface scattering mechanisms. By addressing this gap, we aim to highlight the potential of geometrically engineered nanostructured plasmonic rectifiers.

In this work, we propose a novel mechanism for plasmon-enhanced second harmonic generation through the geometric rectification—*funnelling*—of ballistic electrons. The zero-threshold voltage of ballistic rectifiers, combined with the strong field enhancement from plasmon resonances, allows even low excitation intensities to induce current rectification and second harmonic generation. Additionally, the plasmonic nature of the mechanism provides significant flexibility in tuning the resonance frequency by adjusting the size and shape of the resonators. We will introduce the mechanism and conclude with design guidelines for practical realization

We focus on reflection-symmetric plasmon resonators in the shape of bow ties[12], as shown in Fig. 1, though the principles apply to other geometric rectifiers as well. When photons are irradiated at resonant frequencies, they excite the collective, longitudinal oscillation of conduction electrons, known as the charge transfer mode. Some electrons scatter back from the tapered bow tie walls, creating a *blocking* effect, while others specularly scatter forward, moving through the neck from one wing to the other, producing a *funnelling* effect. This configuration leads to an anharmonic oscillation of the charge transfer mode, facilitating energy transfer to the second harmonic plasmon mode.

To realistically assess nonlinearities in the electron flow, we need to include dynamic effects such as boundary scattering of ballistic electrons and the development of resonant plasmon modes in time. This is done by applying the particle-in-cell (PIC) method[13], which was demonstrated to be a powerful tool for providing insight into the femtosecond electron dynamics in plasmon resonators, including damping and electron scattering effects[14]. PIC naturally models scattering phenomena given its use of discrete particles, allowing us to clearly see the effects of surface scattering very directly. The framework of the PIC technique is presented in the Methods Section.

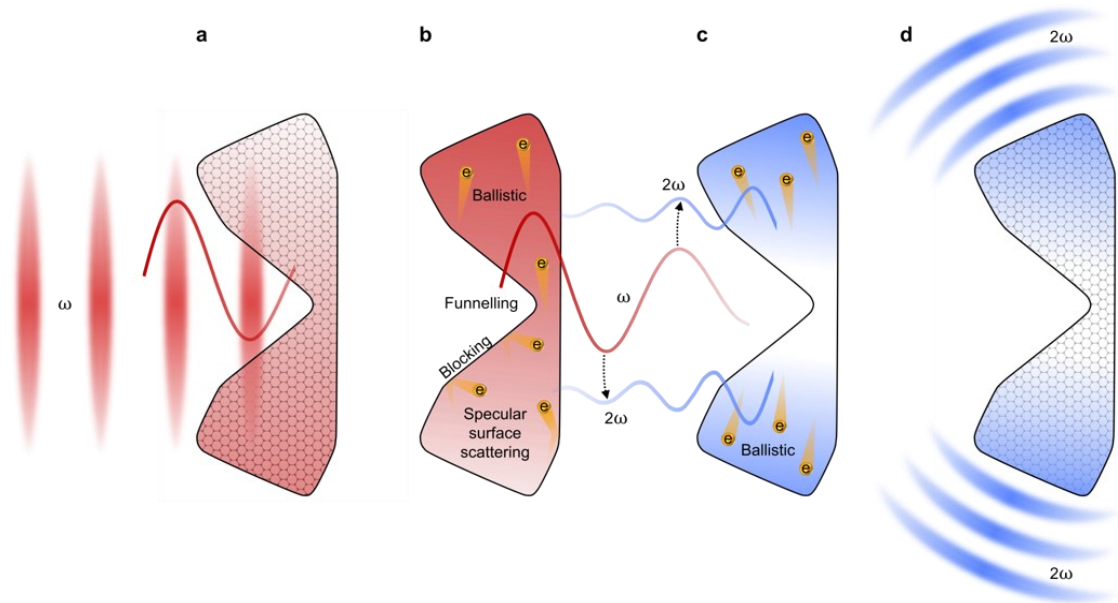

**Fig. 1 | Schematic illustration of nonlinear ballistic electrons funnelling due to specular surface scattering.** Illuminating light with angular frequency $\omega$ excites the first-order charge transfer plasmon, causing ballistic electrons to scatter specularly at the tapered surface and contribute to a second harmonic resonance. This effect enables current rectification and frequency doubling for THz radiation.

## Funnelling-Induced Plasmon Nonlinearity Mechanism

Funnelling-induced plasmon nonlinearity depends on three factors: resonator geometry, electron dynamics, and surface scattering. Each of these is discussed in separate sections.

**Resonator Geometry**

We start the discussion with ballistic electron flow in straight bow tie shaped resonators. Fig. 2a shows schematically how the geometry can be used to control nonlinear electron flow for the charge-transfer plasmon mode in bow tie resonators. The taper angle $\alpha$ can be used to control the ratio between the *blocking* and *funnelling* effects. Both blocking and funnelling effects are present at low taper angle, while the funnelling effect is suppressed at high taper angle due to dominant backscattering. The details for the geometrical designs are shown in Fig. S1 and Fig. S2.

Fig. 2b shows the influence of the taper angle on the response spectrum. The bow tie resonators are 153 nm in length and have an electron density $n_e$ of $10^{26}$ m$^{-3}$. They are illuminated by mid-infrared radiation in the energy range 0.05 – 0.15 eV (12 - 36 THz), covering only the first-order charge-transfer bright mode (**B1**), polarized in the longitudinal direction. We choose to monitor near-field energy to give a fair comparison between bright and dark plasmon modes. Dipolar bright modes such as **B1** can couple to illuminating light, but dark modes do not have that ability, as their resultant field vector is zero, such as for the first dark mode **D1**, in which the electrons simultaneously oscillate in opposite direction between the two wings of the bow tie, as shown in the current density plots of Fig. 2c.

From Fig. 2b we observe that the second harmonic mode **D1** is well-developed only for taper angles between 20 and 30°, due to the occurrence of both *blocking* and *funnelling*. Electron *blocking* results in the broadening of the fundamental mode at large taper angles (SI Section 3), while the combination of *blocking* and *funnelling* results in the generation of second harmonic mode at low taper angle. We note that the effect is only observed for the charge-transfer mode **B1**. The result for the next bright mode of the bow tie structure (**B2**) is presented in Fig. S3, where there is no charge transfer between the two wings across the neck, no broadening of the fundamental mode with increasing taper angle, nor the generation of a second harmonic mode.

The maximum intensity ratio between the fundamental mode and the second harmonic mode in Fig. 2b is achieved at a taper angle of 23°, when there is perfect frequency matching of the **B1** and **D1** plasmon resonance modes: $\omega_{exc} = \omega_{B1} = \frac{1}{2}\omega_{D1}$. The relative resonance frequency of **B1** and **D1** modes can be tuned freely with geometry[15] and will be discussed further below.

In Fig. 2d, we select the geometry with a taper angle of 23° that satisfies the frequency-matching $\omega_{B1} = \frac{1}{2}\omega_{D1}$ and excite the **B1** mode by a narrow-band ($\omega_{B1} \pm 0.001\ eV$), 1000 fs light pulse. The time-dependent intensity of the **B1** mode (black) and second harmonic **D1** mode (purple) are monitored and plotted. Strikingly, Fig. 2d shows that there is a time delay between the **B1** and **D1** peaks. This time delay is a result of *energy transfer* between the fundamental peak ($\omega_{B1}$) and the secondary harmonic ($\omega_{D1}$) via surface scattering (Fig. S6), and it is largest when the modes are closely matched (Fig. S7 and Fig. S8). Therefore, by exciting the low-frequency **B1** mode with light, part of its resonant energy is used to generate a high-frequency second harmonic.

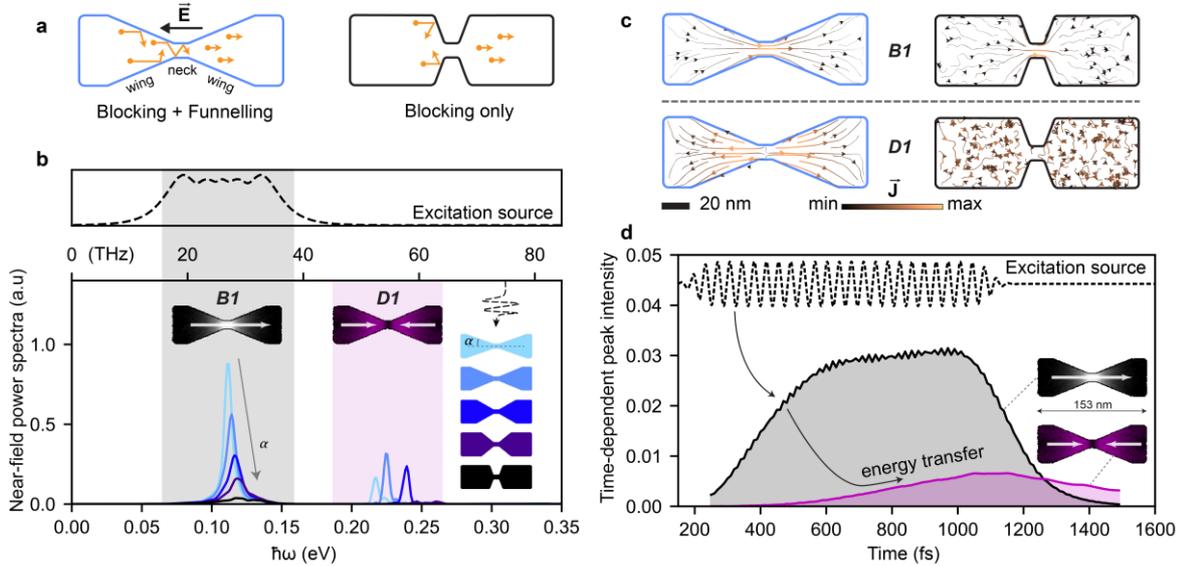

**Fig. 2 | Geometry-dependent nonlinear electron flow due to electron funnelling.** (**a**) Schematic of the angle-dependent response of ballistic plasmons in bow tie structures, showing electron transport for low taper angles (left) and blocked flow for high taper angles (right) due to specular backscattering. (**b**) Near-field energy spectra for bow tie structures with taper angles of 20°, 23°, 30°, 40°, and 70°, excited with a 20 MV/m field in the 12-36 THz range, highlighting the charge-transfer mode **B1**. (**c**) Current density distribution for the charge-transfer bright mode (**B1**) and dark mode (**D1**) at low and high taper angles. (**d**) Time-dependent intensity of the **B1** and **D1** peaks for a 23° taper angle structure excited with a 1000 fs pulse at the **B1** resonance frequency, with the second harmonic coinciding with **D1**. The narrow-band excitation field strength for (**d**) was 5 MV/m.

**Electron Dynamics**

Ballistic electron transport has been studied extensively in two types of electron gases: ballistic electrons in semiconductors that follow a *parabolic* band structure and ballistic electrons in graphene that follow a linear *Dirac* band structure. In these systems, the charge carrier type can be switched between electrons and holes through either electrostatic or chemical doping. In Fig. 3, we compare the plasmon dynamics between (1) parabolic-band electrons and holes, and (2) Dirac-band electrons and holes, focusing on their contribution to the nonlinearity of the generated plasmons. The pusher formalisms for the parabolic band and Dirac electrons in the PIC code are presented in SI Section 6. As expected, the ratio between forward and backward currents is reversed when switching from electron carriers to hole carriers, as seen in Fig. 3a-b and Fig. 3e-f, showing that the asymmetry is indeed due to particle dynamics, distinguishing this from nonlinearity due to electric field inhomogeneity at metal surfaces[16].

Remarkably, the charge oscillation dynamics of parabolic and Dirac-band electrons are opposite, as shown in Fig. 3a and Fig. 3e, indicating a different relative phase between the fundamental mode **B1** and the second harmonic mode **D1**. To explain this, we plot the electron velocity distribution at the maximum forward and backward current, as shown in Fig. 3c, d, g, and h. In this case, the velocity distribution of the holes is plotted for ease of explanation.

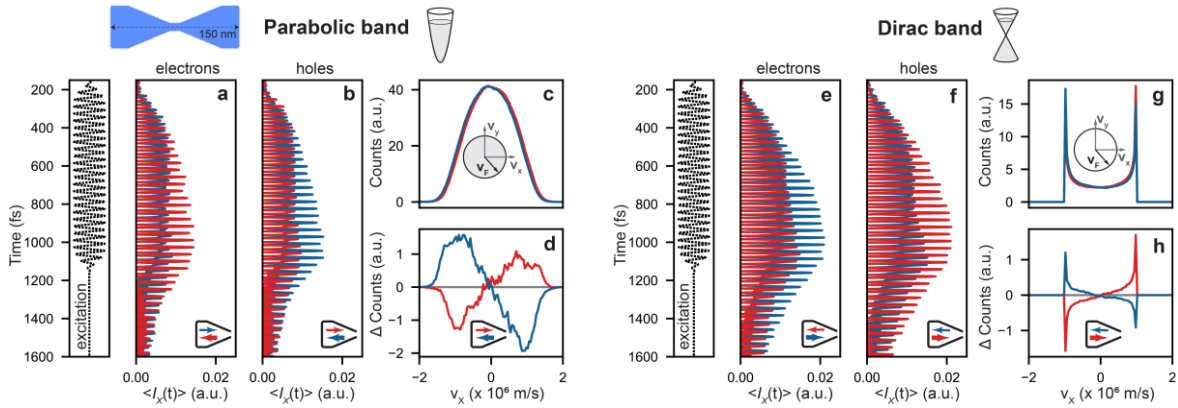

**Fig. 3 | Electron dynamics in parabolic and Dirac-band electrons.** (**a**), (**e**) and (**b**), (**f**) show the forward (red) and backward (blue) current for the left half of the bow tie with a tapering angle of 23°, for electrons and holes as charge carriers, respectively. (**c**), (**g**) Histograms of carrier velocity along the x-direction, with red/blue corresponding to the maximum forward/backward flow. Insets show the velocity distribution in velocity space: (**c**) parabolic carriers have an evenly distributed velocity within the Fermi circle, while (**g**) Dirac carriers have a fixed Fermi velocity. (**d**), (**h**) Histograms of the difference between forward and backward flow at equilibrium, illustrating the opposing trends in (a), (b) vs. (e), (f). The excitation field strength was 5 MV/m.

The unexpected difference between the parabolic band and Dirac charge carriers sheds light on the role of surface scattering in these two systems. Within one-half of the oscillation cycle, a large portion of all the particles undergo surface scattering as the surface scattering rate is comparable with the excitation cycle (Fig. S4). In the parabolic-band system, surface scattering is simply a blocking effect. In the Dirac-band system, however, surface scattering affects the average effective mass of the Dirac carriers. In contrast, Dirac carriers have a directional effective mass (SI Section 6). Surface scattering changes the direction of the charge carriers relative to the applied field, and therefore reduces the effective mass of the Dirac carriers that initially move purely along the longitudinal direction. Enhanced surface scattering in one-half of the oscillation cycle results in larger drift velocity, as shown in our statistical analysis in SI Section 7. We call this effect *surface Dirac anharmonicity* to distinguish it from bulk Dirac anharmonicity[8]. Both are due to the fixed

Fermi velocity of Dirac electrons, but surface Dirac anharmonicity inherently has an even order, while bulk Dirac anharmonicity inherently has an odd order.

**Electron Surface Scattering**

Our PIC simulations being kinetic in nature[13,14] automatically include mechanisms for second harmonic generation: a second-order term due to nonuniformity of the electric field at the surface[16–19], plasmonic drag effects[20,21] and bulk Dirac anharmonicity[8,11]. Electron-surface scattering effects do not play an important role in these mechanisms. They deserve our attention though, as they play a pivotal role in electron funnelling[22–24] and surface Dirac anharmonicity (Fig. S11). To highlight this, we now compare specular and diffuse surface scattering and their effect on second harmonic mode formation. Diffuse scattering results in a weaker fundamental mode for the same excitation intensity due to enhanced damping (Fig. S12a-b). So, instead of using the same excitation intensity for specular and diffuse scattering, we compare the second harmonic peaks for the same intensity of the fundamental mode. In this way, the field enhancements for specular and diffuse scattering are kept the same.

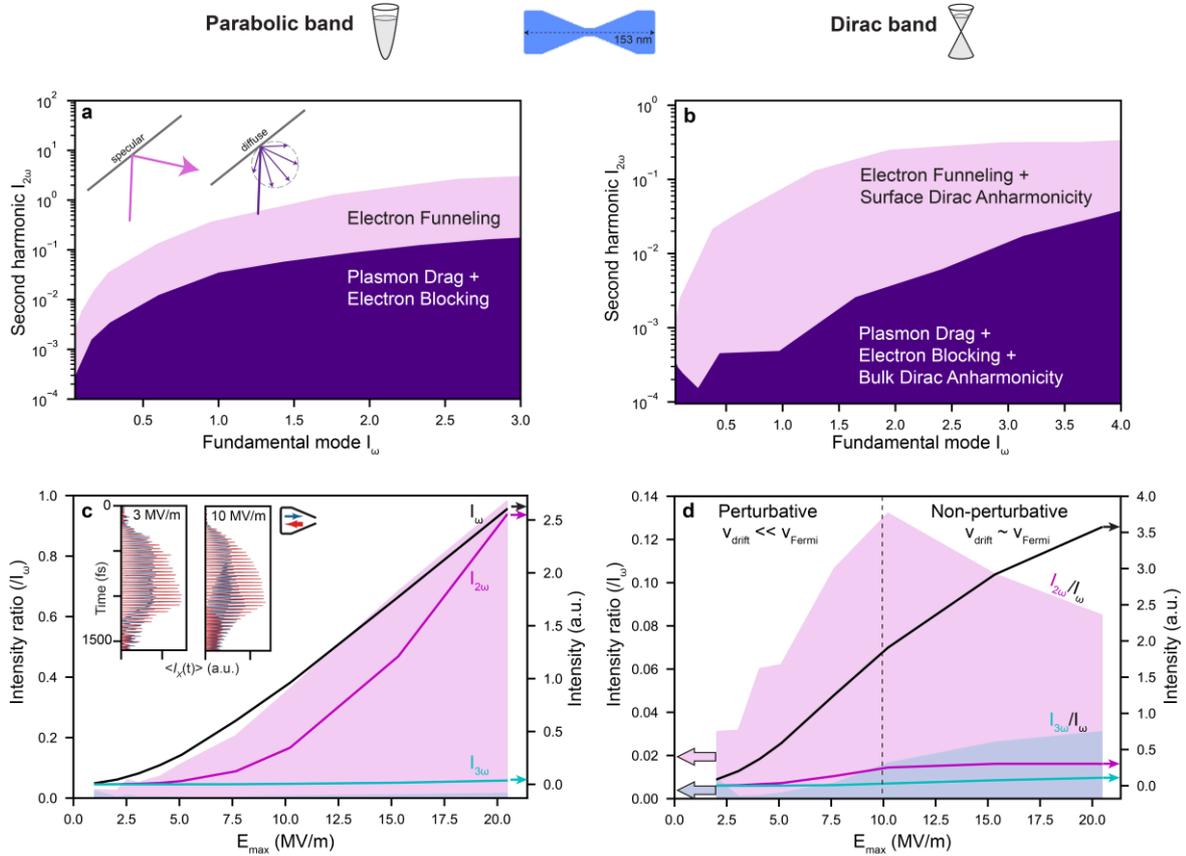

**Fig. 4 | (a-b) Second harmonic generation for different field strengths and scattering mechanisms**. (**a**) and (**b**) Intensity of the second harmonic mode $I_{2\omega}$ versus fundamental mode $I_\omega$ for parabolic-band (**a**) and Dirac-band (**b**) electrons, with specular (pink) and diffuse (purple) surface scattering, for a 153 nm long bow tie with a 23° taper angle. (**c**) and (**d**) Intensity dependence on excitation field strength, showing the fundamental, second, and third harmonic intensities (right axis) and the intensity ratios of the fundamental to the second (pink) and third (cyan) harmonics (left axis) for parabolic-band electrons (**c**) and Dirac-band electrons (**d**), excited by a 1000 fs pulse at the **B1** resonance frequency.

Figures 4a and 4b show that the second harmonic mode for specular scattering is 1-2 orders higher than for diffuse scattering. The difference can be attributed to the contribution of electron funnelling for both parabolic and Dirac band electrons, and to surface Dirac anharmonicity for the case of Dirac band electrons. The results in Fig. 4a-b show that electron funnelling relies on specular electron surface scattering. Specular scattering also enhances surface Dirac anharmonicity, which is otherwise suppressed for diffuse scattering (Fig. S11e). These results therefore highlight the importance of specular surface scattering and surface engineering in designing nonlinear devices based on ballistic electrons, which has not been shown or discussed explicitly in previous work. Previous graphene-based ballistic rectifier designs[3,4] were demonstrated on patterned graphene with rough surfaces and may exhibit almost 100% diffuse scattering[25–28]. Specular scattering is expected for semiconductors such as GaAs and Si where the etched surfaces are smooth, forming effective boundaries for carrier scattering[1,29,30]. Alternatively, a smooth boundary for graphene can be introduced in electrostatically defined geometries[31].

The enhancement of the second harmonic intensity due to electron funnelling and surface Dirac anharmonicity also indicates that these mechanisms can be used to reduce the required incoming field intensity for a second harmonic generation. This is explored in Figs. 4c-d.

Figure 4c shows a monotonous increase of the second harmonic mode with the incoming field for parabolic-band electrons. The insets show the time-domain profile for low and high field intensity; remarkably, the ratio between the forward and backward current does not increase. For low-to-medium incident fields, the time-dependent intensity of the secondary peak does not vary with excitation field strength At large field strengths however, the time delay reduces, which we ascribe to plasmon drag[20,21]; Fig. S13 and Fig. S14 show this in more detail. These observations show that electron funnelling is even expected at low illuminating field intensity, similar to how ballistic rectifiers are expected to operate at low-to-no threshold voltage.

For Dirac-band electrons, we observe two regimes, indicated in Fig. 4d as perturbative and non-perturbative. In the perturbative regime ($v_{\text{drift}} \ll v_F$) where the illuminating field is low, the second harmonic peak intensity ratio increases with excitation field strength. At large field strengths, however, the system enters the non-perturbative regime where $v_{\text{drift}} \sim v_F$, shown in more detail in Figs. S12c-d. In the non-perturbative regime, *bulk* Dirac anharmonicity becomes significant, evidenced by the increase in the third harmonic peak intensity. This further emphasizes the potential use of electron funnelling in the low-field perturbative regime.

The field intensity used in our simulations is in the range of $> 10^9 - 10^{11}\ W/m^2$, for an effective electron mass of $1m_e$. For the case of electrons in graphene with an effective mass $\sim 0.02 - 0.04m_e$ for electron densities in the range of $10^{12} - 10^{13}\ cm^{-2}$, we expect to observe frequency doubling at $10^5 - 10^7\ W/m^2$, a few orders of magnitude lower than the threshold of a non-perturbative regime ($> 10^9 - 10^{10}\ W/m^2$)[10,11,32–34]. The smallest practical field intensity is so low that we do not reach this threshold in our simulations before we are limited by the numerical noise that is inherent in PIC simulations due to discrete particle noise[35]. It is therefore promising that the effect may still be observable at even lower excitation intensity than we can reach, potentially giving a wide practical domain for applications.

# Design Principles

### Frequency Matching Conditions

The results in Figs. 3-4 are for the ideal case of perfect resonance matching conditions $\omega_{B1} = \frac{1}{2}\omega_{D1}$. The frequency ratio of the illumination $\omega_{B1}$ and the second harmonic $\omega_{D1}$ can be tuned by many degrees of freedom in the bowtie design, such as the neck width, neck length, wing length, the periodicity of the array[12] (Fig. S15) or the dispersion relation of the electron gas: 2D vs. 3D electron gas. In this section, we explore the second harmonic generation for conditions where the resonant frequencies do not perfectly match, to establish practical design principles.

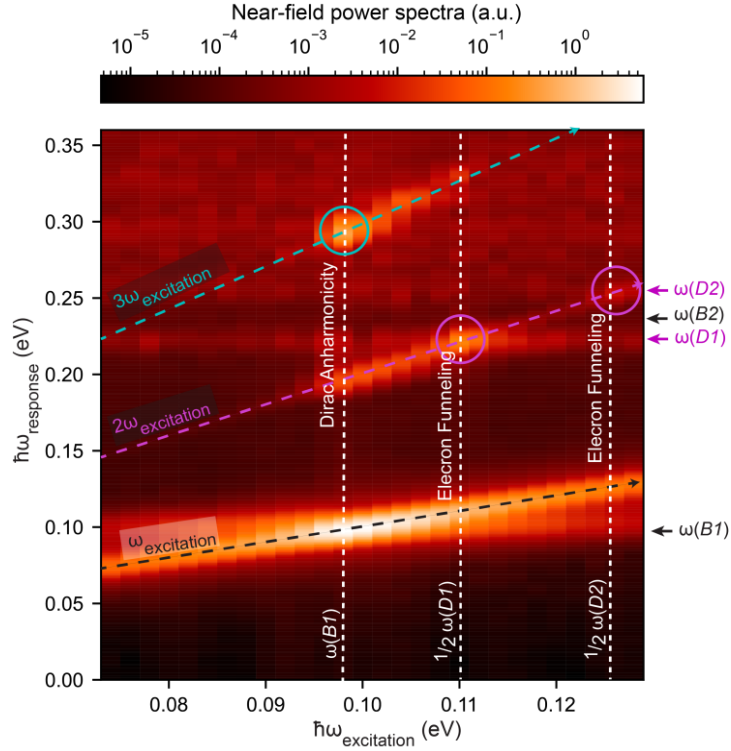

**Fig. 5 | Frequency matching and geometry selection**. Near-field power spectra of the response for varying narrow-band (0.001 eV) excitation frequencies of the same structure as in Fig. 4 with Dirac electrons. The resonance frequencies are detuned (Fig. S15) such that $\omega_{B1} \neq \frac{1}{2}\omega_{D1}$. Guiding lines highlight the fundamental, second, and third harmonic modes, showing the frequency matching conditions for second and third harmonic generation. The excitation field strength was 20 MV/m, in the non-perturbative regime, to highlight bulk Dirac anharmonicity.

Fig. 5 presents a generalized case where the resonance frequencies are detuned such that $\omega_{B1} \neq \frac{1}{2}\omega_{D1}$. It shows the power spectra of the response with varying narrow-band ($\pm 0.001$ eV) excitations for the Dirac electron case. Unlike conventional plasmon-enhanced nonlinearity, the second harmonic intensity is not maximized at the **B1** resonance with the highest field enhancement. Instead, it is globally maximized when the second harmonic frequency $2\omega_{excitation}$ coincides with the dark mode **D1**. Another local maximum is observed at the second plasmon dark mode **D2**. Both plasmon dark modes exhibit an anti-parallel mode profile between the two wings, consistent with the opposite phase of surface scattering; for clarity, the profiles of these modes are given in Fig. S17. No local maximum is observed at the coincidence of $\omega_{excitation}$ with second bright mode **B2**. This is expected, given that the phase of this mode profile does not match with the surface-scattering phase. The third harmonic, however, is maximized only at **B1** resonance as the third harmonic is mainly due to bulk Dirac anharmonicity. The result here serves as another evidence that distinguishes the mechanisms presented in this work (funnelling, blocking, surface Dirac anharmonicity) from conventional plasmon-enhanced nonlinear mechanisms that heavily rely on the enhancement of fundamental field[11,36,37].

**Materials Selection**

In this section, we present the main criteria for geometry-induced nonlinear plasmons: (I) Ballistic transport, where surface scattering is the main scattering mechanism in the system. (II) Critical damping, where the surface scattering rate $\gamma_{\text{surface}}$ is comparable with the oscillation frequency $\omega$. For the first criterion on ballistic transport, we show that the intensity of the second harmonic mode decreases to zero when the inelastic mean-free-path due to impurity scattering becomes more dominant than electron-surface scattering, as shown in Fig. S18. The second criterion is determined by the Fermi velocity and the electron density, i.e. the material from which the plasmon resonator is made. Given a certain resonator geometry, its resonance frequency $\omega$ can be controlled through electron density. A relevant example is that of graphene, for which the electron density can be actively tuned by electrostatic gating or via chemical doping. Conversely, the material of the resonator can be fixed and the geometry can be changed. This has a similar effect on the surface scattering rate: keeping the Fermi velocity $v_F$ the same but changing the characteristic length of the resonator $L$, the surface scattering rate $\gamma_{\text{surface}}$ is affected via $\gamma_{\text{surface}} \sim v_F/L$.

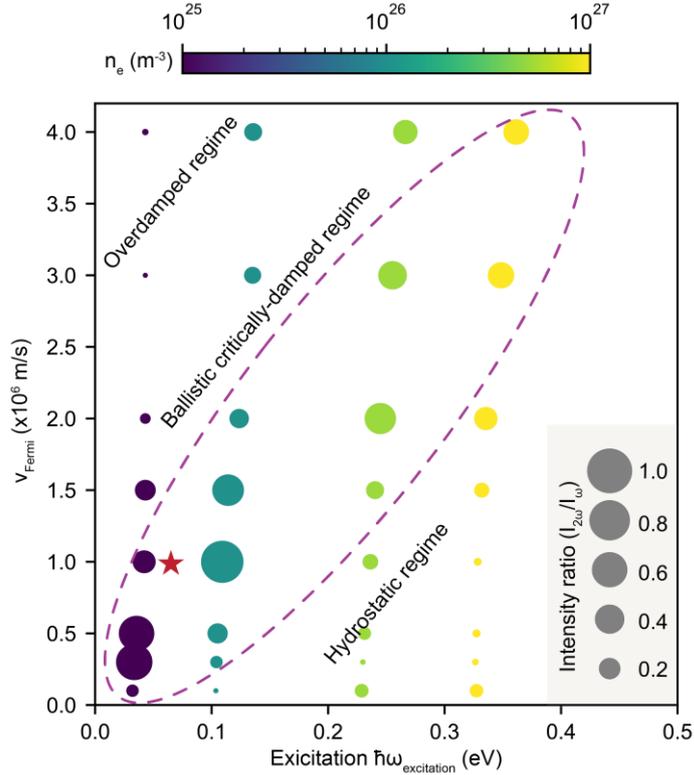

**Fig. 6 | Operating regimes and materials selection.** The same bow tie geometry as in Fig. 4 is excited with $\omega_{\text{excitaion}} = \omega_{B1} = \tfrac{1}{2}\omega_{D1}$. The marker size indicates the ratio between the second harmonic and the fundamental mode intensity as a function of Fermi velocity and excitation frequency. $\omega_{exc}$ is tuned by varying the electron density: for purple, $n_e$ is $10^{25}$ m$^{-3}$, for blue-green $n_e$ is $10^{26}$ m$^{-3}$, for light green $n_e$ is $5\times10^{26}$ m$^{-3}$, and for yellow $n_e$ is $10^{27}$ m$^{-3}$. The excitation field strength is normalized to be proportional to the Fermi velocity such that the ratio $\frac{v_{\text{drift}}}{v_F}$ is consistent while varying $v_F$ for a fair comparison, accounting for the effect of field strength presented in Fig. 4. The above results are from 2D simulations of 3D materials, see also SI Section 12. The operating regime of free-hanging graphene with a sheet density of $10^{13}$ cm$^{-2}$ with the same geometry is indicated by the red star in this figure. More 3D simulation results for graphene are shown in Fig. S17.

These effects are summarized in Fig. 6 in which the relative intensity of the second harmonic $I_{D_1}/I_{B_1}$ is indicated by the size of the markers, for electron densities from $10^{25}$ to $10^{27} m^{-3}$ and Fermi velocities from $1 \times 10^5$ to $4 \times 10^6$ $m/s$. Three regimes can be distinguished: (1) the *hydrostatic regime*, in which the surface scattering rate is comparable with the oscillating frequency. This happens when the Fermi velocity is low, and the electrons are considered frozen compared to the oscillation cycle and are therefore not affected by surface scattering. Plasmons in this regime follow the local response approximation and contribution from individual electron dynamics is negligible[38]. With increasing $v_F$, electrons scatter on the surfaces more often, resulting in a larger difference between forward and backward current, which translates to a larger relative intensity of the secondary peak. This is regime (2), the *ballistic critically-damped regime*. A further increase of $v_F$ broadens the fundamental peak and weakens the secondary harmonic peak (Fig. S19). This is (3) the *overdamped regime*, in which the surface scattering rate is much larger than the oscillating frequency. Within one oscillation cycle, an electron may scatter several times on the surface, causing decoherence to the plasmon oscillation, and the resonance quickly dampens out.

The red star in Figure 6 marks the operating regime of graphene. For this, we performed a more computationally intensive 3D simulation of 2D free-hanging graphene with the same bow tie geometry and sheet density of $10^{13}$ cm$^{-2}$, represented by the red star in this figure (Fig. S17). With this as a starting point, the operating regime for doped graphene can be estimated for various sheet densities via $\omega \propto n_e^{1/4}$. The 153 nm long graphene bow tie has an operating energy range from 0.065 eV for the **B1** mode to 0.078 eV for half the frequency of the **D1** mode. This is well below the corresponding Fermi level of 0.26 eV and therefore avoids interband transitions.

It should be noted that the case we present here is for a semiclassical, non-interacting electron gas. Our simulations only include *intraband* plasmons; the effect of *interband* transitions is not included. The simulations also do not account for quantum effects and the effect of different types of graphene edges. This will be worth considering, given that the zigzag edge results in significantly more plasmon damping than the armchair edge[39]. The critical factor for these devices to work is the specular scattering of electrons at the surfaces. Graphene ballistic rectifiers can exhibit significant specular scattering in the nonlinear transport regime, possibly due to bias-induced trapped charges at the surface[24,40,41]. Alternative ways to introduce smooth barriers in graphene for specular scattering are local dopant profiles[42] and electrostatically-introduced charge barriers[31]. The effect of such electrostatic charge barriers is demonstrated in Fig. S20 with a soft boundary condition where second harmonic generation due to electron funnelling persists.

## Applications: Tunable Upconversion and Rectification

Here we present a practical application for geometry-induced nonlinearity that delivers photon upconversion and current rectification. The second harmonic peak we have been focusing on for the second harmonic generation is a plasmonic dark mode, which couples only weakly with far-field irradiation. By slightly modifying the geometry, as shown in Fig. 7, we can turn the dark mode into a radiative mode that can be measured in the far-field[37,43]. Fig. 7a-c shows how the illumination and emission can have nearly perpendicular polarization by introducing a small break in the bow tie symmetry. The incoming radiation at 26 THz polarized in the x-direction is resonant with the charge transfer plasmon **B1**, which excites the second harmonic **D1** at 52 THz polarized in the y-direction. This design therefore functions as a plasmonic resonator that can be used as a frequency doubler for THz radiation. As the operating frequency of our resonator design strongly depends on its geometry, it provides a highly tuneable platform for photon upconversion in the infrared.

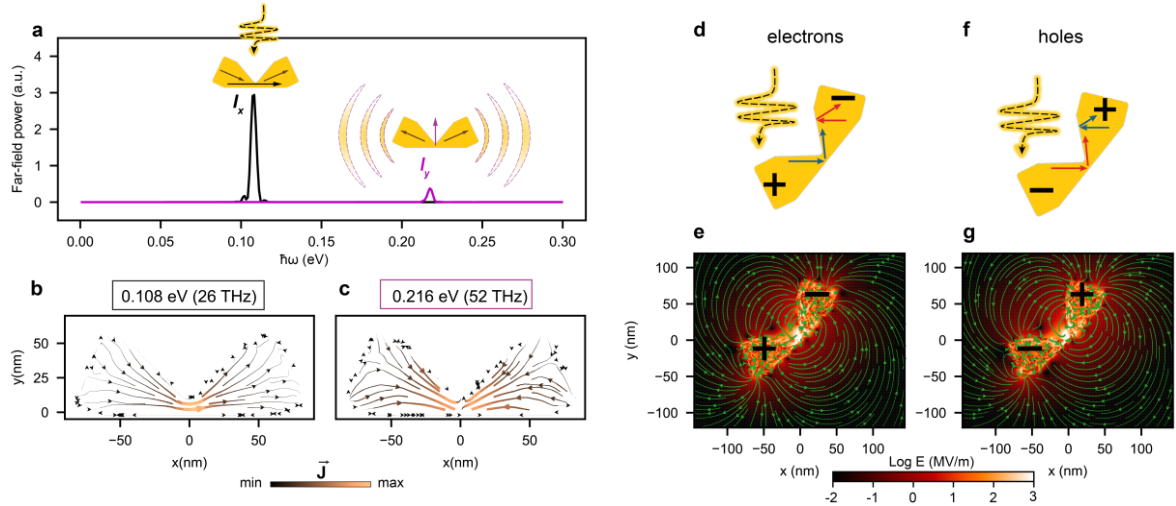

**Fig. 7 | THz generation (a-c) and current rectification (d-g)**. (**a**-**b**) Modified symmetrical bow tie for coupling the dark secondary peak to the far-field. The structure is illuminated at the charge-transfer mode **B1** resonance, polarized along the bow tie length, with a high excitation field (20 MV/m) to highlight charge accumulation. (**a**) Far-field power spectrum of the net dipole current $I_x$ (black) at the fundamental frequency and $I_y$ (pink) at the second harmonic frequency. (**b**) Current density distribution at the fundamental and (**c**) second harmonic frequency. (**d**-**g**) Rectification due to geometry-induced nonlinearity, with symmetry broken by changing the polarization away from the longitudinal axis. (**d**), (**f**) Schematic of rectification for ballistic electrons and holes, respectively. (**e**), (**g**) Averaged DC electric field distribution over the 1000 fs duration, showing electron and hole accumulation in the top bow tie wing.

The nonlinearity discussed in this work arises from surface scattering—specifically electron funnelling and surface Dirac anharmonicity—which differs from anharmonicity caused by (1) nonlocal plasmonic drag and (2) bulk Dirac electron anharmonicity at high illumination intensities. Previously proposed nonlinear plasmonics in graphene required field intensities on the order of $10^9$-$10^{10}$ W/m$^2$ [10,32], which are impractical for ambient infrared light. In contrast, our zero-threshold voltage ballistic rectifiers, as shown in Fig. 7d-g, enable low-field, light-driven second harmonic generation, offering practical applications such as converting ambient infrared radiation into electrical power.

Ballistic electron flow in plasmon resonators offers a promising approach for THz-frequency charge transport[44], enabling practical use of the infrared spectrum. By exploiting resonant modes that couple efficiently with light, we demonstrate how the size and shape of the resonators dictate second harmonic generation and current rectification. Specular surface scattering is key to realizing nonlinear ballistic plasmon resonators, requiring precise control over surface quality and patterning. The robust design rules we have established for optimizing second harmonic generation in the ballistic regime will enable practical applications in tunable photon upconversion, low-field current rectifiers, and energy harvesting from ambient infrared radiation.

## Materials and Methods

**The Particle-in-Cell framework for plasmon simulations**

We simulate the optical response of an electron gas confined within a hard boundary using the Particle-in-Cell (PIC) method using the SMILEI code[45]. The framework for simulating plasmons with PIC has been discussed previously[13,14]. Free electrons are simulated as charged macro-particles moving in a self-consistent electromagnetic field, governed by Maxwell's equations:

$$\begin{cases} \dfrac{\partial \boldsymbol{E}}{\partial t} = \dfrac{1}{\mu_0 \varepsilon_0} \nabla \times \boldsymbol{B} - \dfrac{\boldsymbol{J}}{\varepsilon_0}, \\ \dfrac{\partial \boldsymbol{B}}{\partial t} = -\nabla \times \boldsymbol{E}; \end{cases} \quad (1)$$

and the Newton-Lorentz equation of motion:

$$m_a \frac{d(\gamma_a \boldsymbol{v}_a)}{dt} = e(\boldsymbol{E}_a + \boldsymbol{v}_a \times \boldsymbol{B}_a) \quad (2)$$

Here, $\gamma_a, m_a, q_a$ and $\boldsymbol{v}_a$ are the relativistic factor, mass, charge, and velocity of the charged particle $a$. As the name suggests, PIC divides the space into cells (Yee-grid), similar to conventional finite-difference time-domain (FDTD)[46] field simulations. The electric field $\boldsymbol{E}$ and current density $\boldsymbol{J}$ are calculated at the center of each cell side, the magnetic field $\boldsymbol{B}$ at the center of each face, and the charge density $\rho$ at the nodes. A charged particle $a$ moving across cells will be pushed by the electric and magnetic field $\boldsymbol{E}_a$ and $\boldsymbol{B}_a$ interpolated from the surrounding cells at the position of the particle[45] (second-order interpolation for the shape function of the particle). In return, the position and momentum of the particles are similarly projected at each cell to obtain $\boldsymbol{J}$ and $\rho$ as material-response inputted in Maxwell's equations. The particle pusher for Equation (2) is performed with the Boris scheme[47], which updates the momentum and position of the particles after each time step.

The particles are initialized with a Fermi-Dirac distribution or with a fixed Fermi velocity for parabolic-band and Dirac-band charge carriers, respectively. The simulation cell-size is chosen to be 1×1 nm$^2$ in order to resolve the Debye length of the free electron plasma. Each cell is initialized with 64 macro-particles distributed regularly. The simulation box is 512×512 nm$^2$ (SI Section 11). The incoming light is polarized in the x-direction, the longitudinal axis of the bow ties. The EM boundary condition is set to be periodic for the y-direction and silver-muller[48] for the x-direction. The simulation parameters used for the presented results are consolidated in Table S1 (SI).

**Near-field and far-field power spectra**

The calculated near-field power spectra are estimated from the kinetic induction energy, which is proportional to the integral of oscillating current power across the length of the bow tie[49] $\sim \int (|I_x(\omega, y)|^2 dy$. Instances of peak intensity $I_\omega$ and $I_{2\omega}$ refer to the peak intensities of the corresponding frequencies in the near-field power spectra. The far-field spectra in Fig. 7 are calculated from $\sim |\int \omega I_x(\omega, y) dy|^2$ for the fundamental mode polarized in x-direction and $\sim |\int \omega I_y(\omega, x) dx|^2$ for the second harmonic mode polarized in y-direction[11].

**Two dimensional simulations**

In this work, we performed 2D (3v) simulations for parabolic electrons and 2D (2v) simulations for Dirac electrons to demonstrate fixed Dirac velocity. A comparison between 2D (2v) and 2D (3v) simulations is shown in Fig. S16 showing similar results. Even though the 2D simulations effectively describe a bulk material unbounded in the dimension normal to the simulation space, they still provide relevant information for the case of a 2D electron gas (*e.g.*, doped graphene, which is one of the ideal materials platforms for ballistic electron transport). This allows us to avoid more computationally expensive 3D simulations for this study. As shown in Fig. S17, 2D and 3D simulations have similar plasmon modes and mode profiles for structures relevant for this work. We acknowledge that the dispersion relations in 2D electron gases in doped graphene and 3D electron gases simulated in this work differ, resulting in differences in the relative frequency of the various modes. However, the bow tie shape has a relatively high degree of freedom[12]. The resonant charge transfer mode frequency depends largely on the neck width and length. In contrast, the resonance frequencies of the dipolar modes (plasmon oscillations in the individual wings) depend more strongly on the dimensions of the wings. The screening length is also slightly different between 2D and 3D electron gas, resulting in different magnitudes of nonlocal effects. This is not necessarily a problem; it was already discussed above that nonlocality is not the main driving force for the effect discussed in this work. Besides, the results discussed above also apply to the 3D plasmons in heavily doped semiconductors[50,51], which also operate in the mid-IR/THz frequency range while supporting high carrier mobility.

The dispersion relation can provide another degree of freedom to satisfy frequency matching conditions. In literature, ballistic diodes are usually realized in a Field Effect Transistor (FET) configuration, either with graphene or with semiconductor heterostructures, where the charge carrier density in these systems is usually tuned electrostatically through a metal gate. The metal gate will induce a screening effect that modifies the plasmon dispersion relation from a relation to a linear dispersion[52,53]. In the scope of this work, however, we do not include the effect of a metal gate. Alternatively, free-hanging graphene can also be doped without a metal back gate while supporting plasmons with square-root dispersion relation with high quality factors[54]. On the other hand, other 2D metals, such as metallic transition metal dichalcogenides[55,56], exhibit a flat dispersion relation due to nonlocal electrostatic screening.

**Simulating Dirac electrons**

Parabolic band electrons have a finite effective mass that can be used as particle mass $m_a$ in the conventional Boris pusher scheme. Dirac-band electrons, however, exhibit a zero effective mass and a fixed Fermi velocity due to the linear dispersion relation of the band structure. To incorporate these features of Dirac-band electrons, we use an inverse effective mass tensor with two eigenvalues corresponding to an effective longitudinal mass $m_L^* \to \infty$ and tranverse mass $m_T^* = p_F/2v_F$, resulting in an effective plasmon mass $m_{plasmon}^* = 2m_T^*$ (a detailed derivation can be found in SI Section 6). This essentially means that the Lorentz force purely rotates the velocity direction while the magnitude is kept constant at $v_F$. In PIC, we describe these dynamics with a modified Boris pusher. The Dirac particle is first accelerated in the $\boldsymbol{E}$ and $\boldsymbol{B}$ fields, similar to the conventional Boris pusher with effective mass $m_T^*$. Then, the velocity of the particle is normalized with the Fermi velocity (Fig. S9). This is equivalent to the particle having a transverse mass $m_T^*$ and longitudinal mass $m_L^* \to \infty$. For comparison between parabolic band electrons and Dirac band electrons, we specify the transverse mass of the Dirac band electrons of our artificial material such that its plasmon mass is the same as that of the parabolic band electrons $m_{T,Dirac}^* = \frac{m_{parabolic}^*}{2} = \frac{m_e}{2}$.

**Hard boundary conditions**

The material boundary is specified with a field of surface normal vectors $\boldsymbol{n_S}$ defined at the nodes of the Yee-grid, with non-zero values outside the boundary of the designed structure. At each timestep after the particle pusher, the normal vector at each particle position $\boldsymbol{n_{S,\alpha}}$ is interpolated with the nearest-neighbour interpolation to determine whether surface scattering is applied when $|\boldsymbol{n_{S,\alpha}}| > 0$ and $\boldsymbol{n_{S,\alpha}} \cdot \boldsymbol{p_{\alpha,t}} > 0$.

For specular scattering, its momentum is specularly reflected:

$$\boldsymbol{p_{\alpha,t,f}} = \boldsymbol{p_{\alpha,t}} - 2\frac{\boldsymbol{n_{S,\alpha}} \cdot \boldsymbol{p_{\alpha,t}}}{(|\boldsymbol{n_{S,\alpha}}|)^2}\boldsymbol{n_{S,\alpha}}$$

For diffuse scattering, we conserve the magnitude of the momentum while its direction is randomly chosen to be at an angle $\theta$ with the $\boldsymbol{n_{S,\alpha}}$ with the probability density $\sim cos(\theta)$ using an inverse sampling method.

Its position is updated to keep particles within the boundary as follows:

$$\boldsymbol{x_{\alpha,t,f}} = \boldsymbol{x_{\alpha,t}} - \frac{\boldsymbol{n_{S,\alpha}} \cdot (\boldsymbol{x_{\alpha,t}} - \boldsymbol{x_{\alpha,t-\Delta t}})}{(|\boldsymbol{n_{S,\alpha}}|)^2}\boldsymbol{n_{S,\alpha}}$$

# Acknowledgments

This research is supported by the Ministry of Education, Singapore, under its AcRF Tier 2 (MOE-T2EP50122-0016) (H.T.B.D., M.B.), and MOE-T2EP50223-0001 (H.T.B.D., W.L., M.B.). We also acknowledge support from the Singapore University of Technology and Design Kickstarter Initiative SKI 2021-02-14 (G.N., W.L.) and SKI 2021-04-12 (W.L.), and the National Research Foundation, Singapore, via Grant No. NRF-CRP26-2021-0004 (G.N., W.L.). The authors would like to thank Dr. Zackaria Mahfoud (Institute of Materials Research and Engineering, A*STAR) and Dr. Ding Wenjun (Institute of High Performance Computing, A*STAR) for their mentorship and fruitful discussions.

# Author contributions

H.T.B.D., W.L., and M.B. conceived the research. H.T.B.D. developed the methods and performed numerical simulations. H.T.B.D. prepared the original draft, which was edited and reviewed by G.N, W.L., and M.B. The project was supervised by W.L. and M.B.

# Competing interests

The authors declare no competing interests.

# Supplementary Information for

**Nonlinear Terahertz Resonances from Ballistic Electron Funnelling**

Hue T.B. Do *et al.*

*Corresponding authors. Email: lin_wu@sutd.edu.sg, msemb@nus.edu.sg

# Contents

### 1. Asymmetric versus Symmetric bowties:

The symmetrical bow tie shape is chosen to study in this work instead of the asymmetrical bow tie established in transport measurements[1,2] because it supports strong charge-transfer resonances through the neck. In contrast, the asymmetric bow tie has weaker plasmon resonances for the charge-transfer mode, as shown in Fig. S1 below. At the same time, the inversion symmetry is still broken for each half of the bow tie (*wing*), which results in scattering-induced nonlinearity. By starting from the straight bow tie structure, we will also be able to make a convincing case for current rectification via geometric symmetry breaking, by rotating the polarization of the incoming radiation away from the longitudinal direction.

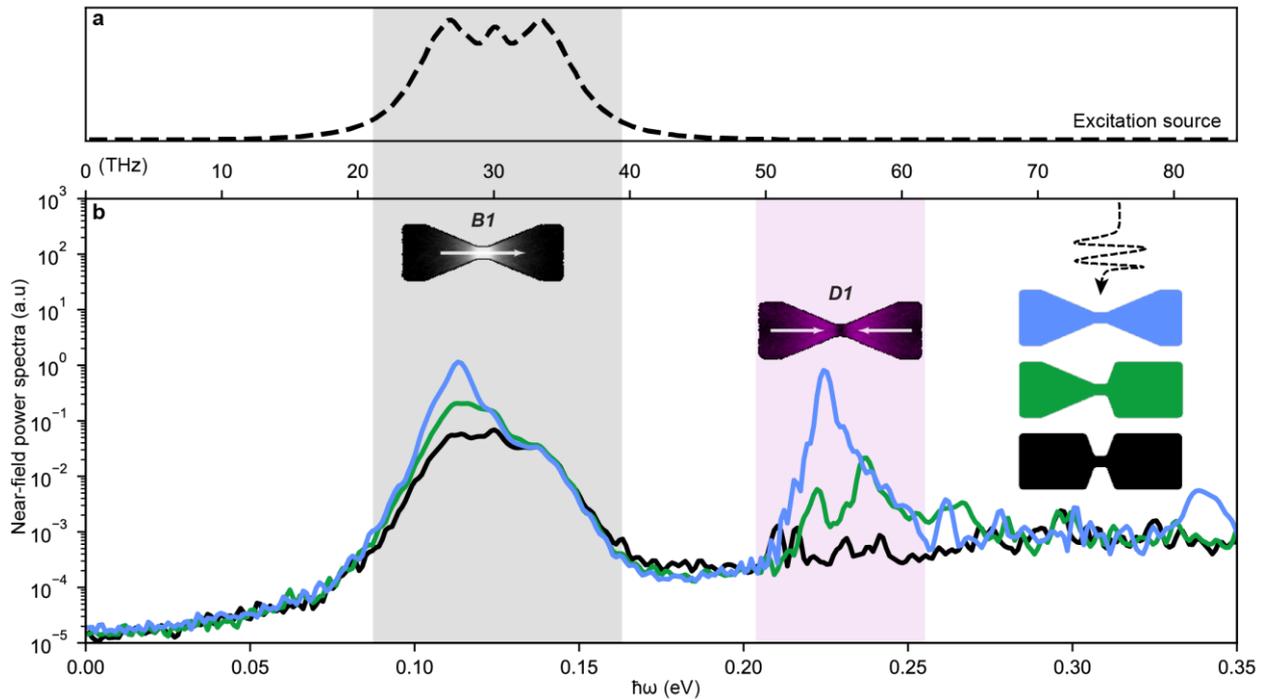

Fig. S1. Comparison between 160 nm long symmetric bowties: The blue structure has a taper angle 23 degrees while the black structure has a taper angle of 70 degrees. The asymmetric bowtie (green) has taper angles of 23 and 70 degrees, and is similar in design to the geometric diode by [2] and [1]. The structures are excited by polarized light covering a frequency window that only includes the *B1* mode (a). Their near-field power spectra are shown in panel b. The asymmetric bowtie also shows the generation of a secondary peak at the *D1* mode profile due to the small taper angle on the left, but with a broadened *B1* peak width and lowered *D1* peak intensity compared to the blue symmetric bowtie due to the blocking effect of the large taper angle on the right.

## 2. Detailed geometry with dimensions

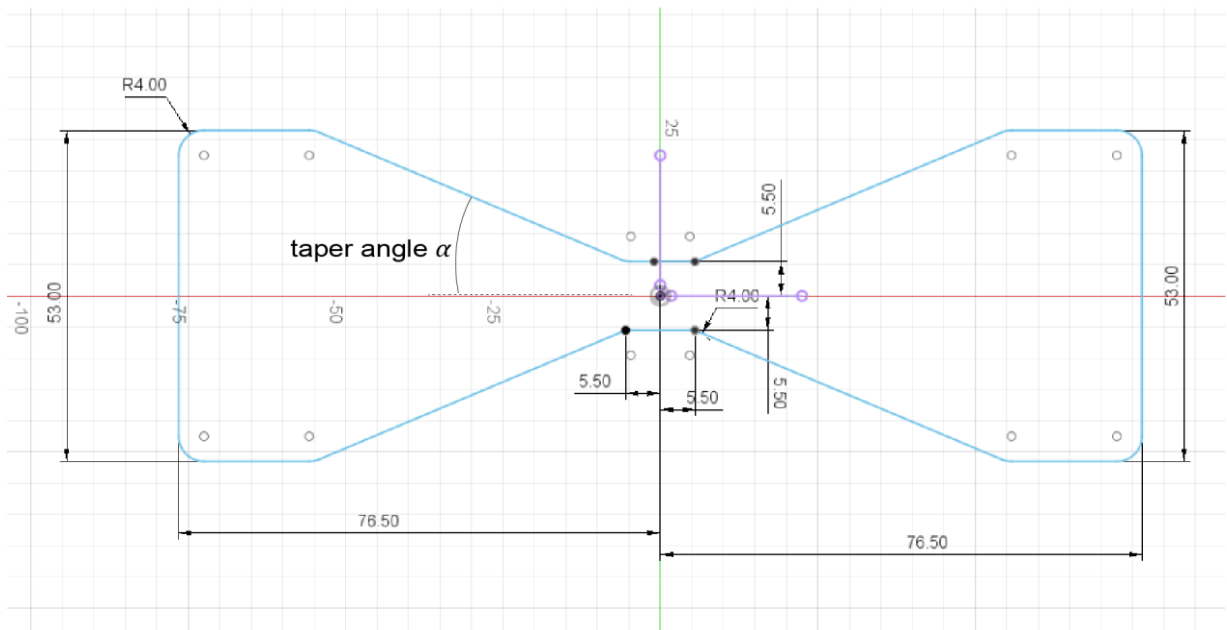

Fig. S2. Geometrical dimensions of the bow tie designs in Fig. 2. Unit: nm

### 3. Discussion on Landau damping

In this study, we assume that the inelastic mean free path of ballistic electrons is much larger than the size of the bow tie resonator. As a result, the damping of the plasmons mainly takes place through radiative damping and electron-surface scattering; the latter is usually referred to as Landau damping. The overall length of the structures here is kept constant, so radiative damping is not expected to vary much with the taper angle. Therefore, we attribute the abnormal broadening of the **B1** peak in Fig. 2b to electron-surface scattering. We extracted the surface scattering rate with varying taper angles in Fig. S4. Here, due to the larger total volume, the electron surface scattering rate has a slight positive relation with the taper angle. This explains the trend in the **B2** mode in Fig. S3, which does not transfer charge between the two wings of the bow ties. The **B1** mode, however, is formed by electrons moving between the two wings across the neck, which are mostly blocked due to scattering on the edges with high taper angle. Conversely, a smaller taper angle allows electrons scattering off a tapered edge to still participate in the charge-transfer mode. In this latter case, the electrons scatter forward rather than backward. As will be discussed below, our simulations cover the thermal regime of the electron plasma where the characteristic length of the geometry $L$ is comparable with $\frac{v_F}{\omega}$. In this regime, Landau damping cannot be naively treated phenomenologically through an effective surface scattering rate defined by the structure size[3]. A detailed discussion on the difference in Landau damping between modes **B1** and **B2** is presented in Fig. S5, clearly showing the effect of the small taper angle on "funnelling" electrons through the neck area.

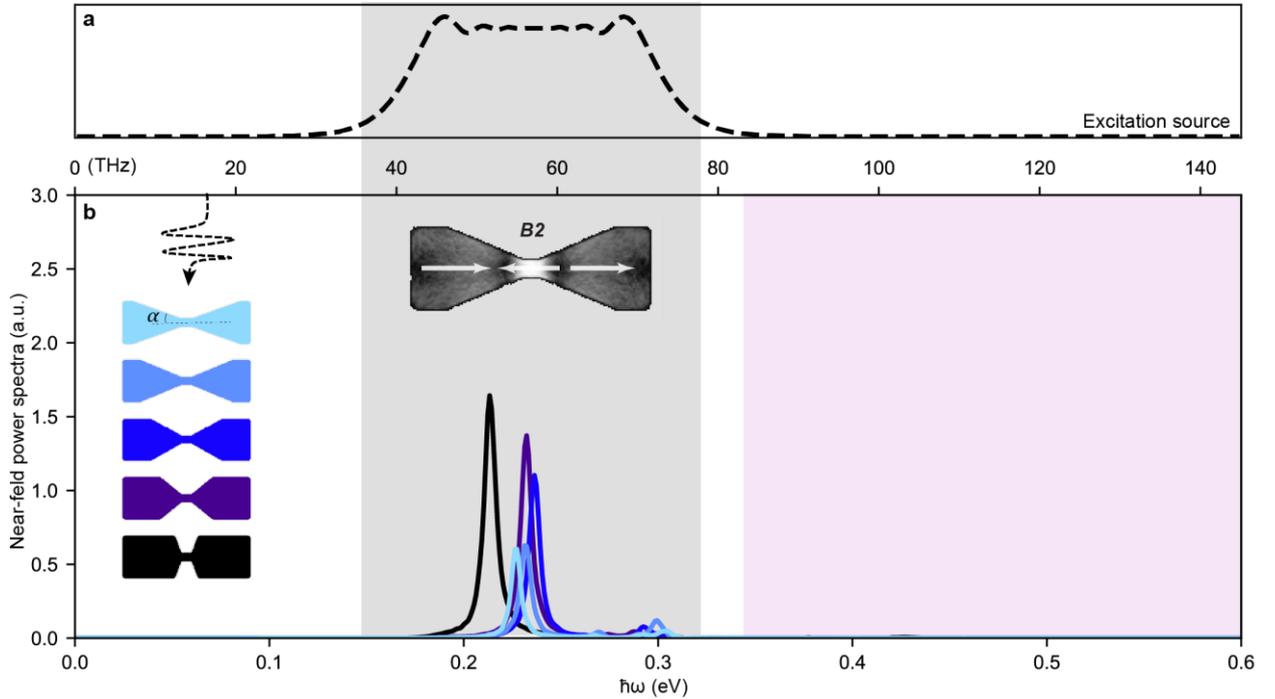

Fig. S3. Near-field power spectra of bowties excited in the frequency window 0.15-0.3 eV (a) covering the second bright mode **B2** (inset of b), corresponding to the parallel dipolar mode where electrons in the two wings oscillate in the same direction (similar to the **B1** mode) but without charge transferring between the two wings (in contrast to the **B1** mode). (b) The spectral response (similar to Fig. 2b) shows no significant second harmonic generation (pink boxed region). There is also no excitation of the *D1* mode, in contrast to the situation in Fig. 2b when the bowties were excited at the charge-transfer mode.

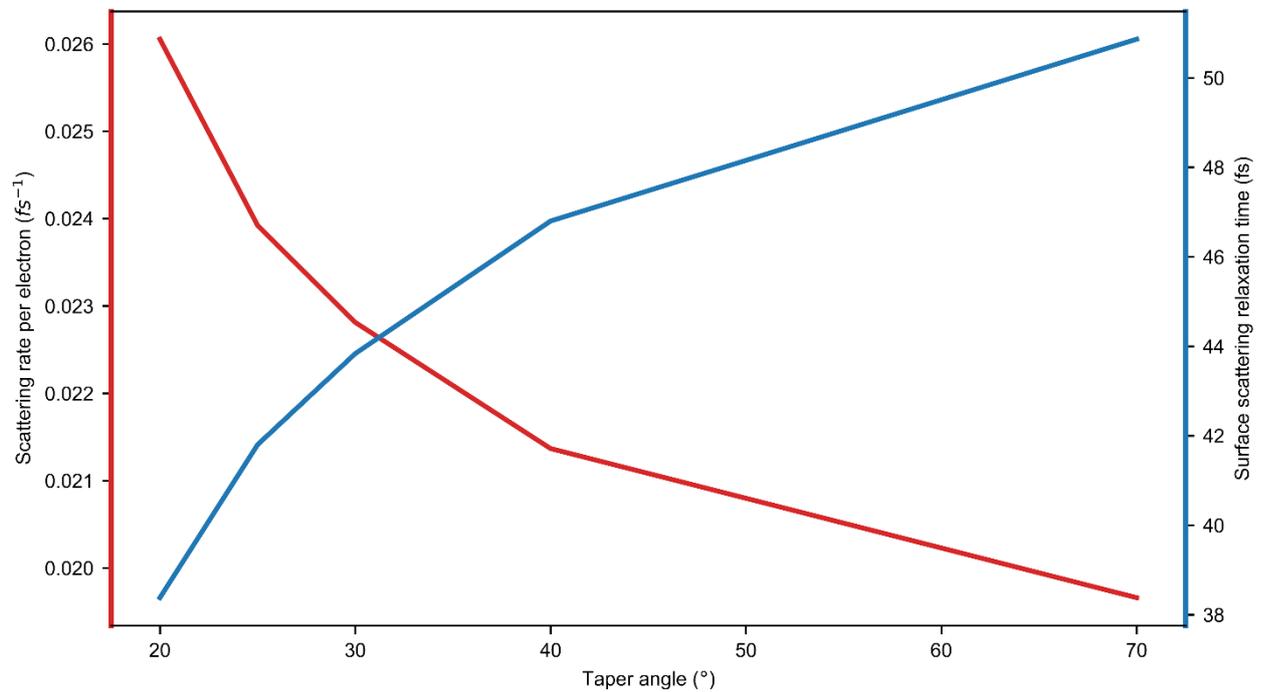

Fig. S4. Scattering rate (left y-axis) and scattering relaxation time (right y-axis) for electron-surface scattering extracted from PIC simulations of bowties with varying taper angles, referring to the structures presented in the inset of Fig. 2b, in the main text. The scattering rate is high for bowties with small taper angles due to their higher surface-to-volume ratio.

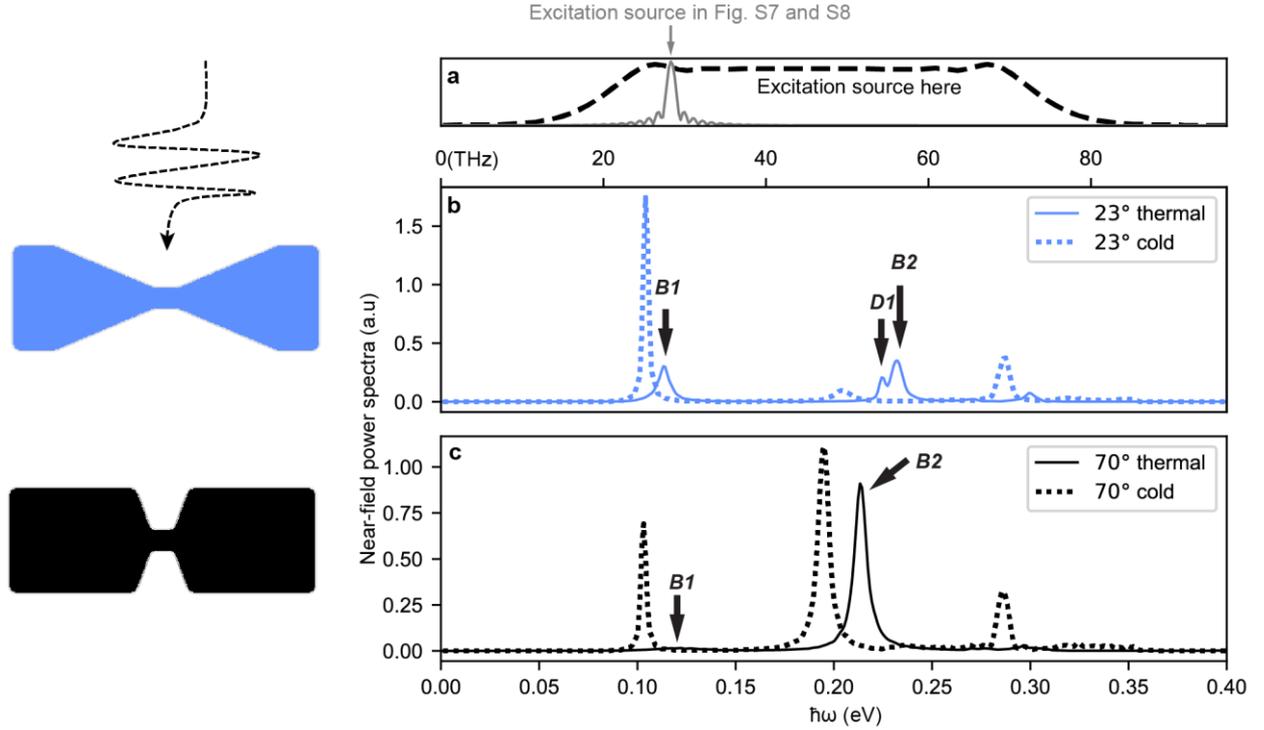

Fig. S5. Comparison between the cold regime and the thermal regime of the bowtie structures (with taper angle 23° and 70°), referring to the main text Fig. 2b (for parabolic-band electrons). (**a**) The structure is excited with a broadband excitation covering both the **B1** and **B2** modes. (**b**) and (**c**) shows the spectral response of the low-taper-angle and the high-taper-angle structure, respectively. The cold regime in the legend refers to an electron gas with Fermi velocity $v_F = 10^5 m/s$, such that the characteristic length of the structure $L \gg \frac{v_F}{\omega}$. As a result, the electron-surface scattering rate is insignificant compared to the plasmon frequency. The thermal regime refers to $v_F = 10^6 m/s$ where $L \sim \frac{v_F}{\omega}$. Here, the electron-surface scattering rate is comparable with the plasmon frequency. In the cold regime, the response is similar to the local response approximation, depending only on the local dielectric function and geometry. In the thermal regime, however, electron-surface scattering becomes significant. Here we observe that the effect of electron-surface scattering when moving from the cold to the thermal regime is more significant for the charge-transfer mode **B1**, where the peak gets weaker and broadened, especially for the high taper angle (70°). For the **B1** mode, the large taper angle blocks most electrons moving between the two wings, explaining the large difference between the thermal and cold cases. Lowering the taper angle allows more electrons to scatter off the tapered edge to the other wing, resulting in more charge-transfer to occur. For the bright dipolar mode **B2**, there is no charge transfer between the two wings. Instead, the electrons accumulate on the two sides of the neck. The difference in intensity of the **B2** peak between the thermal and cold cases is less significant. There is still a large shift in frequency due to enhanced nonlocal effects for the thermal case as the hydrodynamic velocity is close to the Fermi velocity.

## 4. Time-domain dynamics

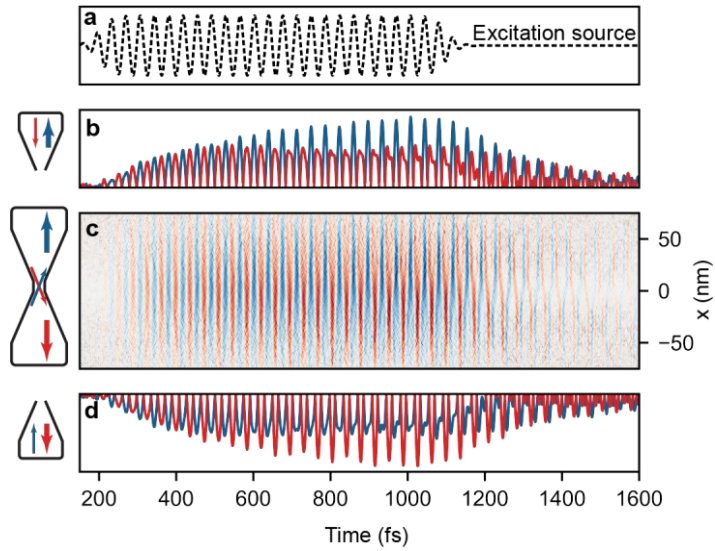

Fig. S6. Time-domain dynamics for the simulation presented in Fig. 2d, showing difference between forward and backward current on each wing of the bow tie. (**a**) Excitation source in time-domain. (**b-d**) Difference between forward and backward currents on each side of the symmetrical bow tie. (**c**) 2D map showing the time-dependent current distribution across the length of the bow tie. Red and blue signify electrons moving in opposite directions. (**b**), (**d**) Sum of the current distribution in (**c**) for $x > 0$ and $x < 0$, respectively. The time-domain dynamics here explains the time-delay between the fundamental and second harmonic mode in Fig. 2d. As shown in panel b and d, in the first 400 fs, there is no difference between the forward and backward current (red and blue). Intuitively, the time-delay is related to the difference in transient time between the forward and backward electron flow as it reaches saturation value. The forward flow is formed by the electrons traveling across the whole length of the bow tie, while the backward flow is formed by electrons traveling about half of the length of the bow tie, as they are blocked by the tapered edge. This results in a longer transient time for forward flow than backward flow.

## 5. Time-domain dynamics for varying frequency matching conditions

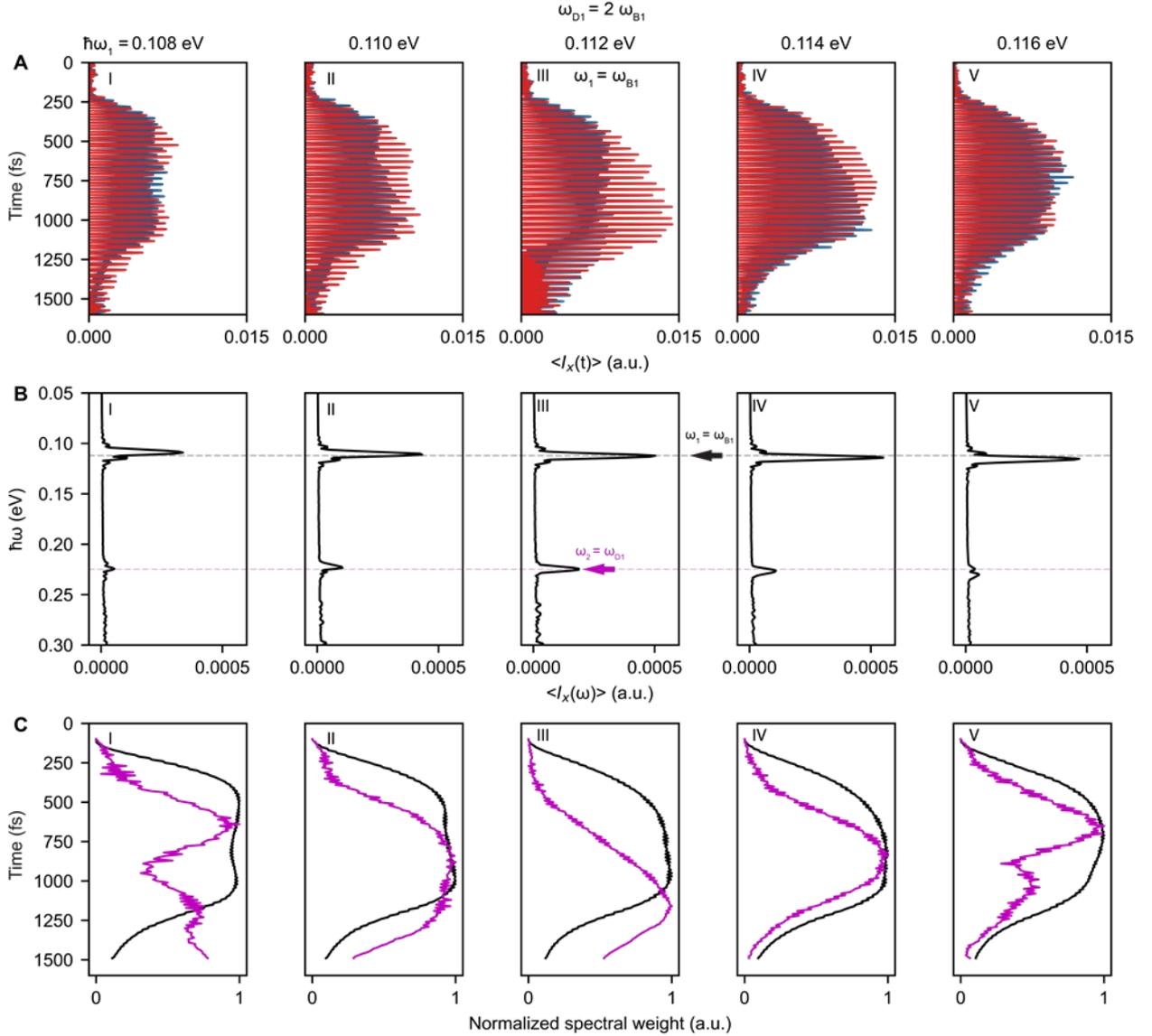

Fig. S7. Narrow-band excitation at taper angle = 23 degrees, when the resonance frequency of the **D1** mode coincides with the second harmonic of the **B1** mode $\omega_{D1} = 2\omega_{B1}$ (for parabolic electrons). The narrow-band excitation source is shown in the previous Fig. S5a in grey, with the central frequency varied around $\omega_{B1}$ as shown on top of columns from (I) to (V), with column (III) showing the case for $\omega_1 = \omega_{B1} = \omega_{D1}/2$. (**a**) Time-dependent oscillating current on the left wing of the bow tie (tapering angle of 23°) with blue and red indicating forward and backward electron flow. (**b**) shows the same oscillating current in frequency domain. (**c**) shows the time-dependent spectral weight of the fundamental peak ($\omega$) and the secondary harmonic peak ($2\omega$). Here the intensity of the secondary harmonic peak and the time delay between the two modes is largest when $\omega_1 = \omega_{B1} = \omega_{D1}/2$ in column (III).

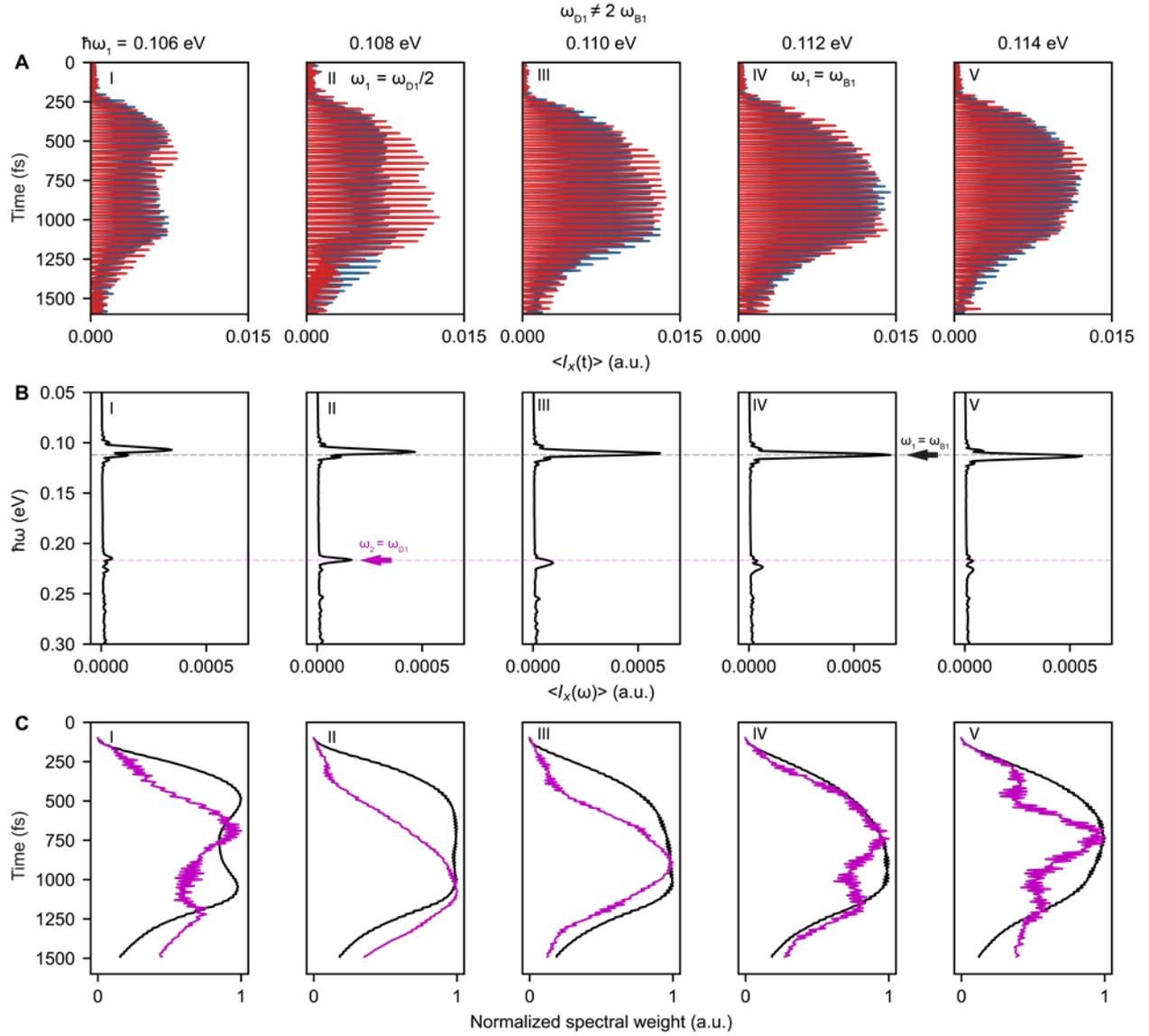

Fig. S8. Narrow-band excitation at taper angle = 20 degrees, when the resonance frequency of the **D1** mode does not coincide with the second harmonic of the **B1** mode $\omega_{D1} \neq 2\omega_{B1}$ (for parabolic electrons). The narrow-band excitation source is shown in Fig. S5a in grey, with the central frequency varied around $\omega_{D1}/2$ and $\omega_{B1}$ as shown on top of columns from (I) to (V), with column (II) and (IV) showing the case for $\omega_1 = \omega_{D1}/2$ and $\omega_1 = \omega_{B1}$, respectively. (**a**) Time-dependent oscillating current on the left wing of the bow tie (tapering angle of 23°) with blue and red indicating forward and backward electron flow. (**b**) shows the same oscillating current in frequency domain. (**c**) shows the time-dependent spectral weight of the fundamental peak ($\omega$) and the secondary harmonic peak ($2\omega$). Here the intensity of the secondary harmonic peak and the time delay between the two modes is largest when $\omega_1 = \omega_{D1}/2$ in column (III), while they are insignificant when the resonance of **B1** is matched $\omega_1 = \omega_{B1}$.

### 6. Dirac band implementation and benchmarking

As graphene is the prime candidate for ballistic transport phenomena, we present here the framework to simulate massless Dirac electrons in graphene with Particle-in-Cell (PIC) simulation. We will then benchmark our PIC results with analytical results for the case of a graphene nano-disk in the nonretarded local approximation.

In PIC, a charged species is described through its mass $m$ and charge $e$ according to the Newton-Lorentz equation of motion, which is applicable for conduction electrons that follow a parabolic band structure $\epsilon_q \propto q^2$, with the effective mass defined as

$$m^{*-1} \equiv \frac{1}{\hbar^2}\left(\partial_q^2 \epsilon_q\right) \tag{S1}$$

Conduction electrons (or holes) in doped graphene follow a linear dispersion relation around the Dirac points $\epsilon_q = \pm \hbar v_F q$, for which the definition of effective mass above returns 0, which disallows the usual use of the Newton-Lorentz equation. The conduction electrons instead are much more readily understood using the Dirac equation of massless relativistic particles with a constant characteristic speed, $v_F \sim 10^6 m/s$.

To incorporate the features of massless Dirac electrons in the Particle-in-cell framework, we adopt an intuitive description based on an inverse effective mass tensor[4], similar to previously described[5]:

$$\overline{m}^{*-1} = \frac{1}{\hbar^2}\nabla_q\nabla_q\epsilon_q(\boldsymbol{q}) = \frac{1}{\hbar^2}\begin{pmatrix} \partial_{q_x}^2 & \partial_{q_x}\partial_{q_y} \\ \partial_{q_y}\partial_{q_x} & \partial_{q_y}^2 \end{pmatrix}\hbar v_F\sqrt{q_x^2 + q_y^2} \tag{S2}$$

The tensor $\overline{m}^{*-1}$ in equation (S2) connects the change in momentum and velocity $d\boldsymbol{v} = \overline{m}^{*-1}d\boldsymbol{p}$. Solving for the eigenvalues and eigenvectors of (S2) yields two eigenvalues $\left\{0, \frac{v_F}{p}\right\}$ corresponding to two eigenvectors $\left\{\binom{p_x}{p_y}, \binom{p_y}{-p_x}\right\}$ that represent the momentum changes in the direction parallel to $\boldsymbol{p}$ (longitudinal) and the direction perpendicular to $\boldsymbol{p}$ (transverse). This means that the effective mass $m_L^* \to \infty$, Dirac electrons moving along the direction of the applied electric field will not accelerate ($d\boldsymbol{v} = 0$), as the velocity of Dirac electrons are capped at the Fermi velocity $v_F$. On the other hand, Dirac electrons moving transverse to the electric field will have an effective mass $m_T^* = p/v_F$. In PIC, we describe these dynamics with a modified Boris pusher as in Fig. S9. The Dirac particle is first accelerated in the $\boldsymbol{E}$ and $\boldsymbol{B}$ fields, similarly to the conventional Boris pusher (Fig. S9 a-c) with effective mass $m_T^*$. Then the velocity of the particle is normalized with the Fermi velocity (Fig. S9d). This is equivalent to the particle having a transverse mass $m_T^*$ and longitudinal mass $m_L^* \to \infty$ as shown in (f) and (g).

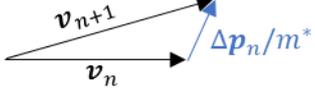

Fig. S9. Modified Boris pusher scheme for parabolic-band electrons and Dirac electrons.

The inverse transverse effective mass $m_T^{*-1}$ varies with momentum $p$ (or wave vector $q$). For simplicity, we may use an inverse effective mass averaged over the whole conduction band of graphene $\langle m_T^{*-1} \rangle$ (in the case of electron-doping) for all the electrons such that the summed current density is conserved as below:

$$\langle \Delta p_T \rangle_{2\pi} \int_0^{p_F} m_T^{*-1}(p)\rho(p)dp = \langle \Delta p_T \rangle_{2\pi} \langle m_T^{*-1} \rangle \int_0^{p_F} \rho(p)dp, \qquad (S3)$$

with

$$\langle \Delta p_T \rangle_{2\pi} = \frac{\int_0^{2\pi} \Delta p_T(\theta)d\theta}{\int_0^{2\pi} d\theta} = \frac{\Delta p}{2}, \qquad (S4)$$

given that the averaged transverse change of momentum over all directions and $\rho(p) \propto p$ is the electron distribution with $p$, and $\Delta p$ is the momentum change from the EM field. We can now obtain:

$$\langle m_T^{*-1} \rangle = \frac{2v_F}{p_F}. \qquad (S5)$$

We can calculate the drift velocity as follows:

$$\langle v_{drift} \rangle = \frac{\langle \Delta p_T \rangle_{2\pi}}{\langle m_T^{*-1} \rangle} = \frac{\Delta p}{p_F/v_F}. \qquad (S6)$$

This yields the plasmon mass per electron $m_{plasmon} = \frac{\Delta p}{\langle v_{drift} \rangle} = p_F / v_F$ as formulated in[6] derived from the kinetic induction energy per electron $E_{kinetic} = \frac{\Delta p^2}{2 m_{plasmon}}$.

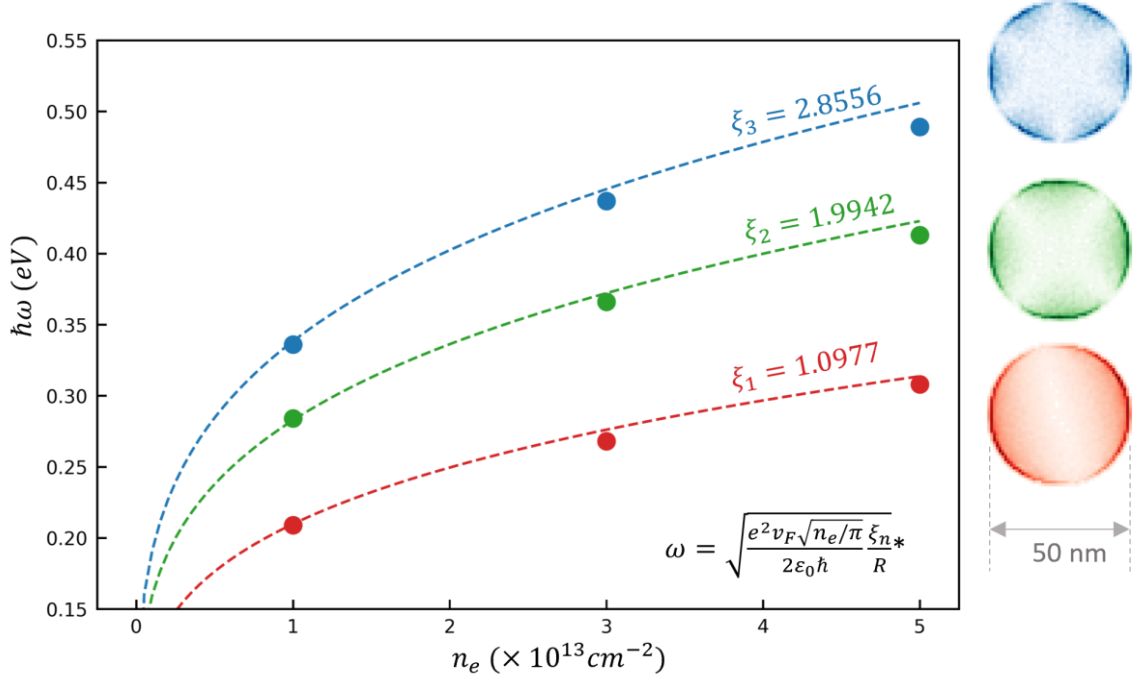

Fig. S10. Benchmarking between PIC simulations and an analytical model for a graphene nano-disk of diameter 50 nm. Here we plot the resonance frequency of three plasmon modes, the charge distribution of which is shown on the right in red, green and blue. The resonance frequency of each mode is plotted in the corresponding colour. The scatter plot shows the PIC results. The dashed lines are plotted based on an analytical model (*Christensen *et al.*[7]) with $\xi_n$ values taken from numerical integration, and are not fitted.

Applying the framework above, we simulate graphene nano-disks with diameters of 50 nm and benchmark this with an analytical model within the nonretarded local-response approximation. The scatter plot in Fig. S10 shows the plasmon energy of three plasmon modes of the graphene nano-disk with varying electron density, which match well with the analytical model (dashed line) presented in *Christensen *et al.*[7]. In this analytical model, the intraband plasmon resonance frequency is established as

$$\omega = \sqrt{\frac{e^2 v_F \sqrt{n_e/\pi}}{2\varepsilon_0 \hbar}} \times \frac{\xi_n}{R}, \tag{S7}$$

with $\xi_n$ being the normalized eigenvalues of the corresponding mode as shown in Fig. S10.

### 7. Surface Dirac anharmonicity

For Dirac electrons, the effective mass is finite along the transverse direction and infinite along the longitudinal direction compared to its velocity. Therefore, the effective plasmon mass of the electron gas depends on its velocity distribution in momentum space. The plasmon mass along the direction of applied field is:

$$m_{plasmon\,x} = \frac{\Delta p_x}{\langle v_{drift}\rangle_x} = \frac{\int_0^{2\pi} n(\theta)d\theta}{\int_0^{2\pi} m_x^*(\theta)^{-1} n(\theta)d\theta} \qquad (S8)$$

With $m_x^*(\theta) = \frac{p_F}{2v_F}\sin^2\theta$ being the effective mass of individual electrons along the direction of applied field.

For a uniform momentum distribution $n(\theta)$, the plasmon mass per electron $m_{plasmon\,x}$ coincides with the cyclotron mass

$$m_{plasmon\,x} = m_c = \frac{p_F}{v_F} \qquad (S9)$$

We show in Fig. S11 below how the plasmon mass of a drifting Dirac electron gas change after a scattering event depending on the drifting direction and the type of surface scattering. Consider a Dirac electron gas with Fermi velocity $v_F$ with uniform initial momentum distribution $n(\theta)$, then accelerated with $v_{drift,\,nominal} = \frac{\Delta p_x}{\langle m_T^*\rangle} = -eE_x\Delta t$ along the $x$ direction, scattering with a taper wall with a taper angle $\alpha$. Scattering with the surface can modify the momentum distribution $n(\theta)$ locally (Fig. S11a-c), hence the plasmon mass along the applied field direction $m_{plasmon\,x}$. Forward scattering, when electrons are accelerated toward the taper wall ($\Delta p_x > 0$) results in a reduction of plasmon mass $m_{plasmon\,x}$ while backward scattering ($\Delta p_x < 0$) results in an increment of the effective mass (Fig. S11d). This results in a cascading effect where reduction of plasmon mass results in larger $v_{drift}$, which further reduces the plasmon mass in forward scattering compared with backward scattering case. The result in Fig. S11d explains the ratio between forward and backward current in Figs. 2e-h and inherently consists of even order harmonics. We call this Surface Dirac anharmonicity as it is the result of surface scattering on directional effective mass of Dirac electrons. Such surface scattering induced anharmonicity does not exist in parabolic band electrons with a scalar electron effective mass. The change in effective mass also depends on the taper angle, with no change is observed for $\alpha = 0\,and\,90°$. In the entire confined bowtie structures at charge-transfer mode **B1**, we expect largest surface Dirac anharmonicity at the taper walls near the centre of the bowties where the accelerating field is the largest.

Fig. S11e shows the ratio of plasmon mass after a forward versus backward scattering event for $v_{drift,\,nominal} = \pm 0.1 v_F$ for specular versus diffuse scattering. As specular and diffuse scattering result in different momentum distribution, specular scattering results in larger surface Dirac anharmonicity than diffuse scattering, as discussed in Fig. 4.

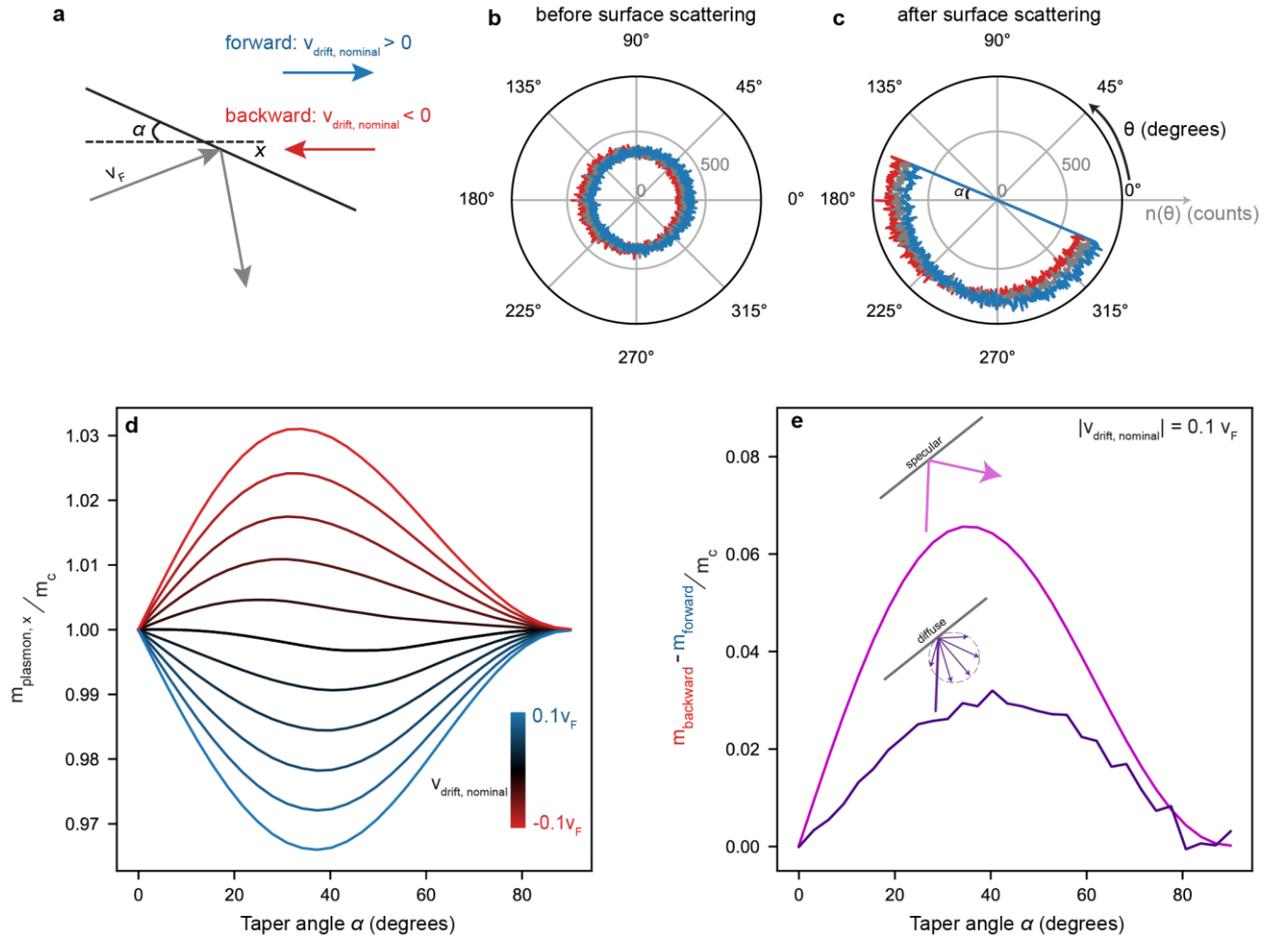

Fig. S11. (**a-c**) Angle-dependent momentum distribution of scattered Dirac electrons before (**b**) and after (**c**) scattering events for a nominal drift velocity $v_{drift,nominal} = \frac{\Delta p_x}{\langle m_T^* \rangle} = -eE_x\Delta t$ along x-direction being $-0.1\ v_F$ (red), 0 (grey) and $0.1\ v_F$ (blue). (**d**) Plasmon mass of specularly scattered electrons with varying $v_{drift,nominal}$ and taper angle $\alpha$. (**e**) Difference between forward and backward plasmon mass for specular (pink) and diffuse scatter (purple) for an electron gas with a nominal drift velocity $v_{drift,nominal} = 0.1v_F$ with varying taper angle $\alpha$.

## 8. Varying excitation field strength

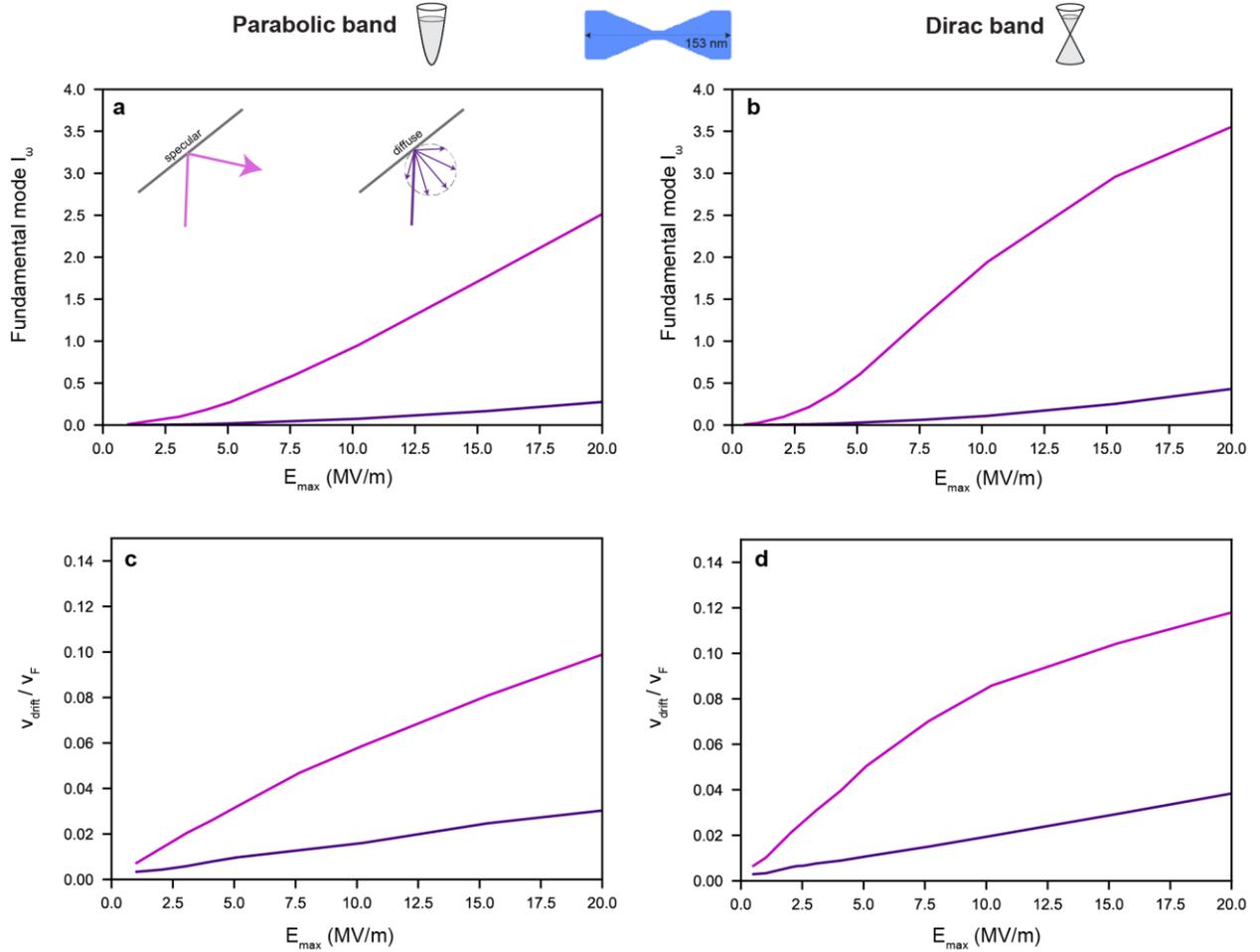

Fig. S12. Comparison between specular surface scattering (pink) and diffuse surface scattering (purple) for parabolic-band electrons (**a-c**) and Dirac-band electrons (**b-d**) for the data presented in Fig. 4. (**a-b**) Intensity of the fundamental mode and (**c-d**) maximum drift velocity versus the excitation field strength for the bow tie structure with taper angle of 23°.

## 9. Time-domain dynamics for the results in Fig. 4.

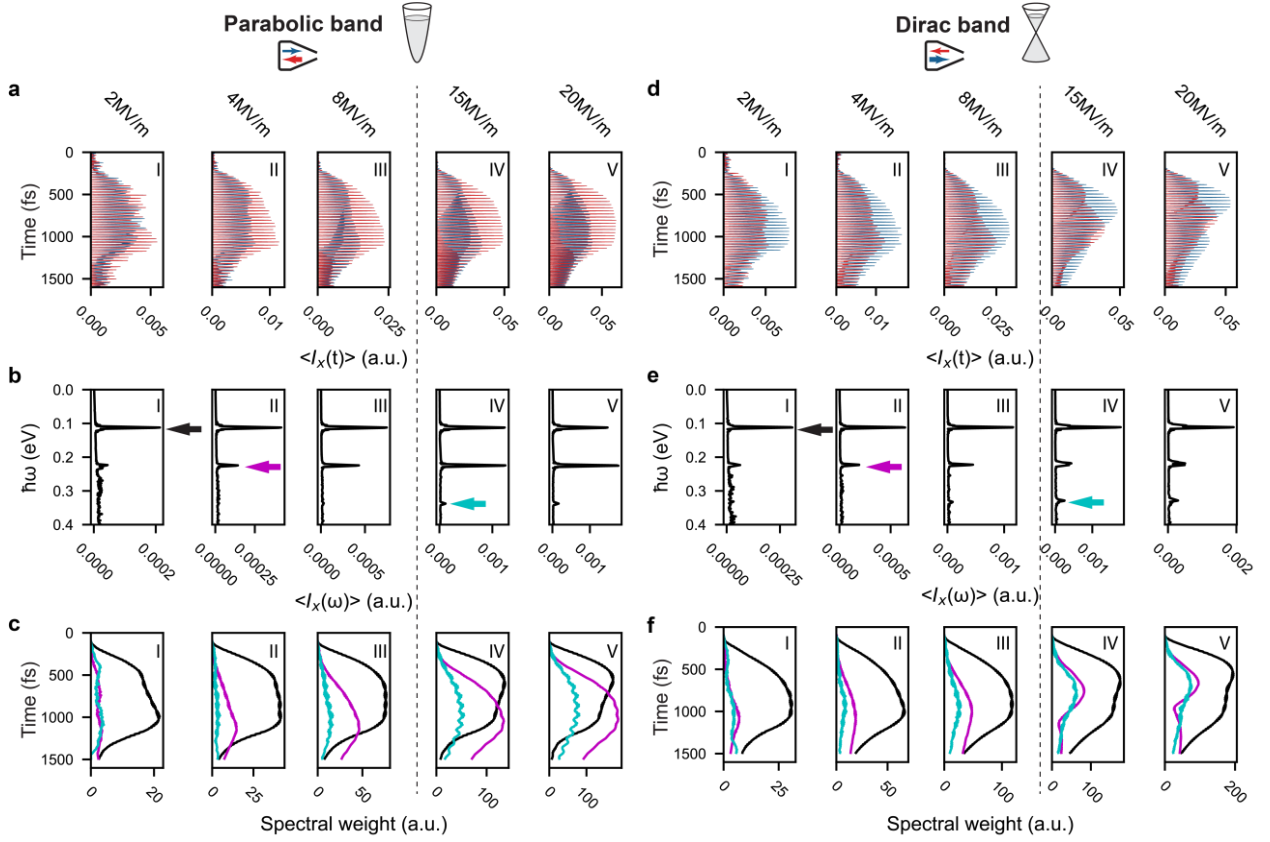

Fig. S13. Intensity-dependent nonlinearity of parabolic-band electrons (**a-c**) and Dirac-band electrons (**d-f**). The electric field amplitude of the incident light at resonance frequency was set to 2 MV/m (I), 4 MV/m (II), 8 MV/m (III), 15 MV/m (IV) and 20 MV/m (V). (**a**), (**d**) Time-dependent oscillating current on the left wing of the bow tie (tapering angle of 23°) with blue and red indicating forward and backward electron flow, as indicated in the top diagram. (**b**), (**e**) FT of panels (**a**) and (**d**) with 3 peaks indicated by coloured arrows: Black is the fundamental peak for the excitation frequency $\omega$, here $\omega = \omega_{B1} = \omega_{D1}/2$, Pink represents the second harmonic at $2\omega$, and cyan the third harmonic at $3\omega$. (**c**), (**f**) Time-dependent spectral weights of the three peaks of the corresponding colour, taken from the FT of a moving 200 fs time-window.

## 10. Time-delay in perturbative and nonperturbative regime

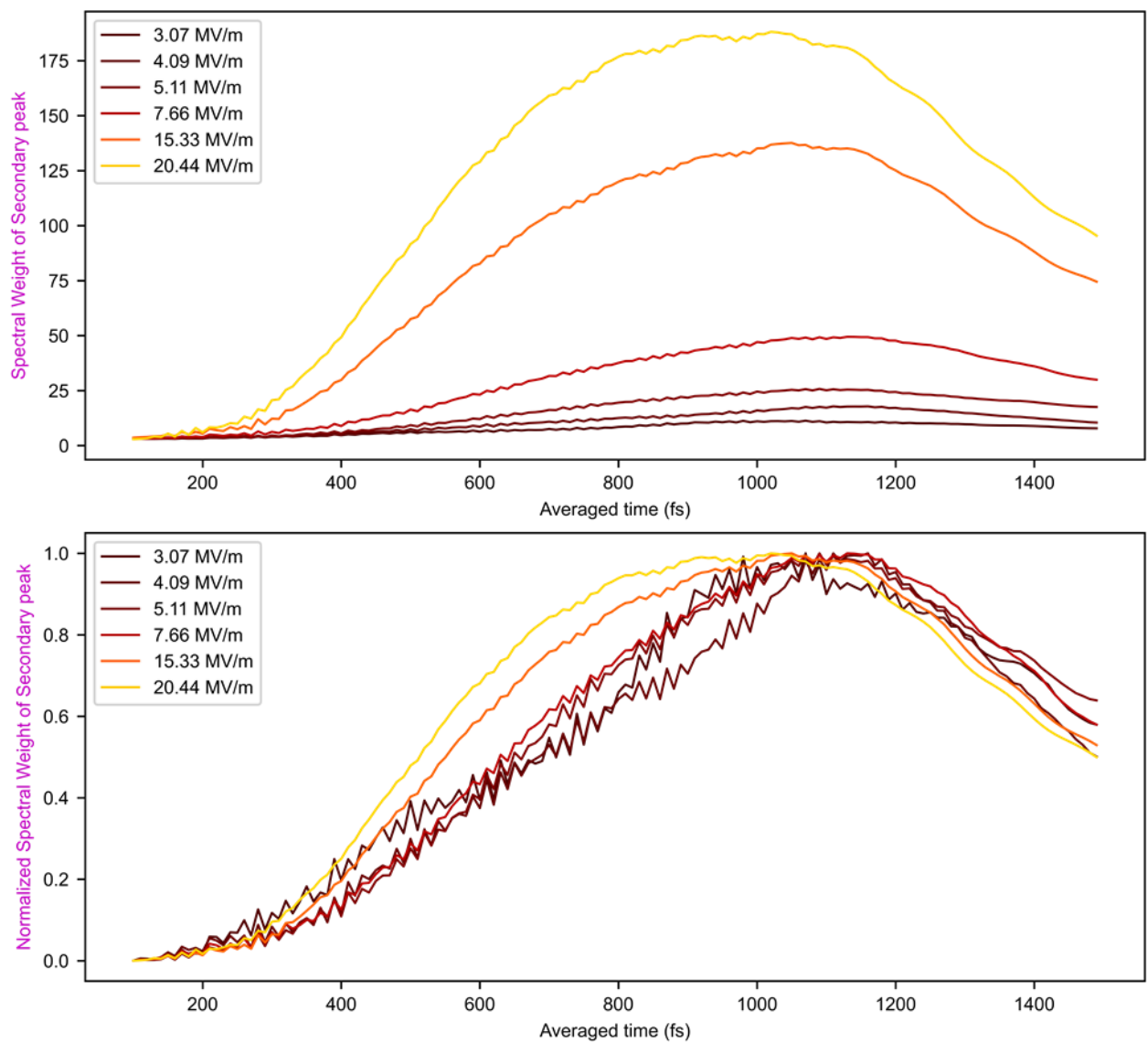

Fig. S14. Time-dependent spectral weight of the secondary harmonic peak shown in Fig. S13c for varying electric field amplitude of the incident light. Top: actual values; bottom: normalized spectral weights.

## 11. Simulation box convergence and array effect:

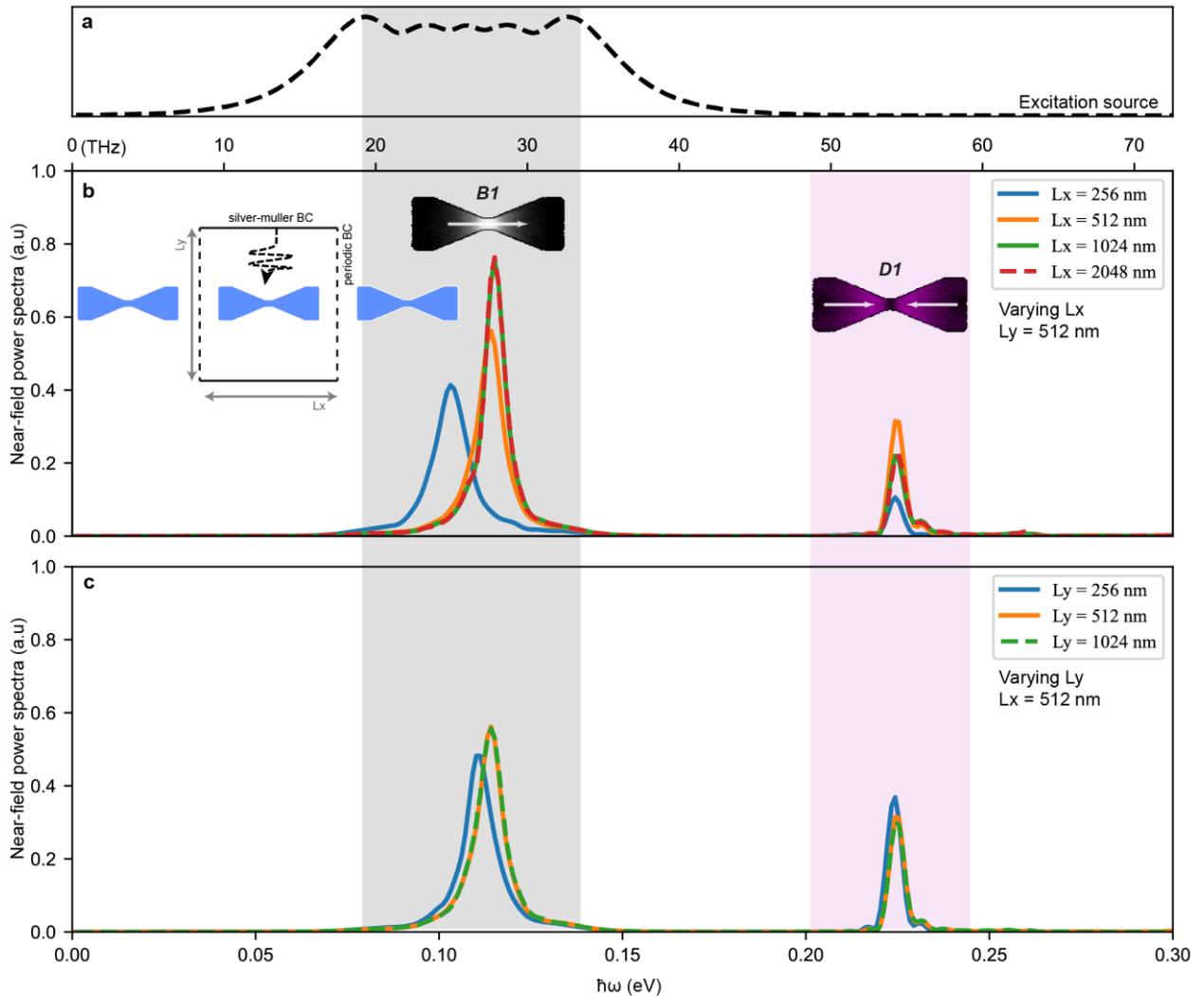

Fig. S15. Simulation box length convergence. The boundary conditions (BC) for the EM field is set to a periodic BC for the x-direction and silver-muller BC (open) for the y-direction. The y-direction boundaries are for incoming and outcoming laser with wavevector $k$ perpendicular to the boundary. A broadband excitation (**a**) is used to excite the bow tie array (inset of **b**) with (**b**) varying array spacing (equivalent to the x-direction box length Lx) and with (**c**) varying y-direction box length Ly. Ly converges with Ly > 512nm and Ly = 512 nm is chosen for all simulations. Lx can be used to tune the relative resonance frequency between **B1** and **D1** modes without changing the bow tie geometry. Lx = 512 nm is chosen for all simulations, except for the results in Fig. 5 where Lx = 256 nm to showcase an off-resonance case.

## 12. 2D and 3D simulations

In this work, we performed 2D (3v) simulations for parabolic electrons and 2D (2v) simulations for Dirac electrons to demonstrate fixed Dirac velocity. A comparison between 2D (2v) and 2D (3v) simulations is shown in Fig. S16 showing similar results. Even though the 2D simulations effectively describe a bulk material, they still provide relevant information for the case of a 2D electron gas (*e.g.*, doped graphene, which is one of the ideal materials platforms for ballistic electron transport). This allows us to avoid more computationally expensive 3D simulations for this study. As shown in Fig. S17, 2D and 3D simulations have similar plasmon modes and mode profiles for structures relevant for this work. We acknowledge that the dispersion relations in 2D electron gases in doped graphene and 3D electron gases simulated in this work differ, resulting in differences in the relative frequency of the various modes. However, the bow tie shape has a relatively high degree of freedom[8]. The resonant charge transfer mode frequency depends largely on the neck width and length. In contrast, the resonance frequencies of the dipolar modes (plasmon oscillations in the individual wings) depend more strongly on the dimensions of the wings. The screening length is also slightly different between 2D and 3D electron gas, resulting in different magnitudes of nonlocal effects. This is not necessarily a problem; it was already discussed above that nonlocality is not the main driving force for the effect discussed in this work. Besides, the results discussed above also apply to the 3D plasmons in heavily doped semiconductors[9,10], which also operate in the mid-IR/THz frequency range while supporting high carrier mobility.

In literature, ballistic diodes are usually realized in a Field Effect Transistor (FET) configuration, either with graphene or with a semiconductor heterostructures, where the charge carrier density in these systems is usually tuned electrostatically through a metal gate. The metal gate will induce a screening effect that modifies the plasmon dispersion relation from a relation to a linear dispersion[11,12]. In the scope of this work however, we do not include the effect of a metal gate. Alternatively, graphene can also be doped without a metal back gate while supporting plasmons with high quality factors[13]. Once doped, the resonance frequency can be tuned passively by controlling the geometry.

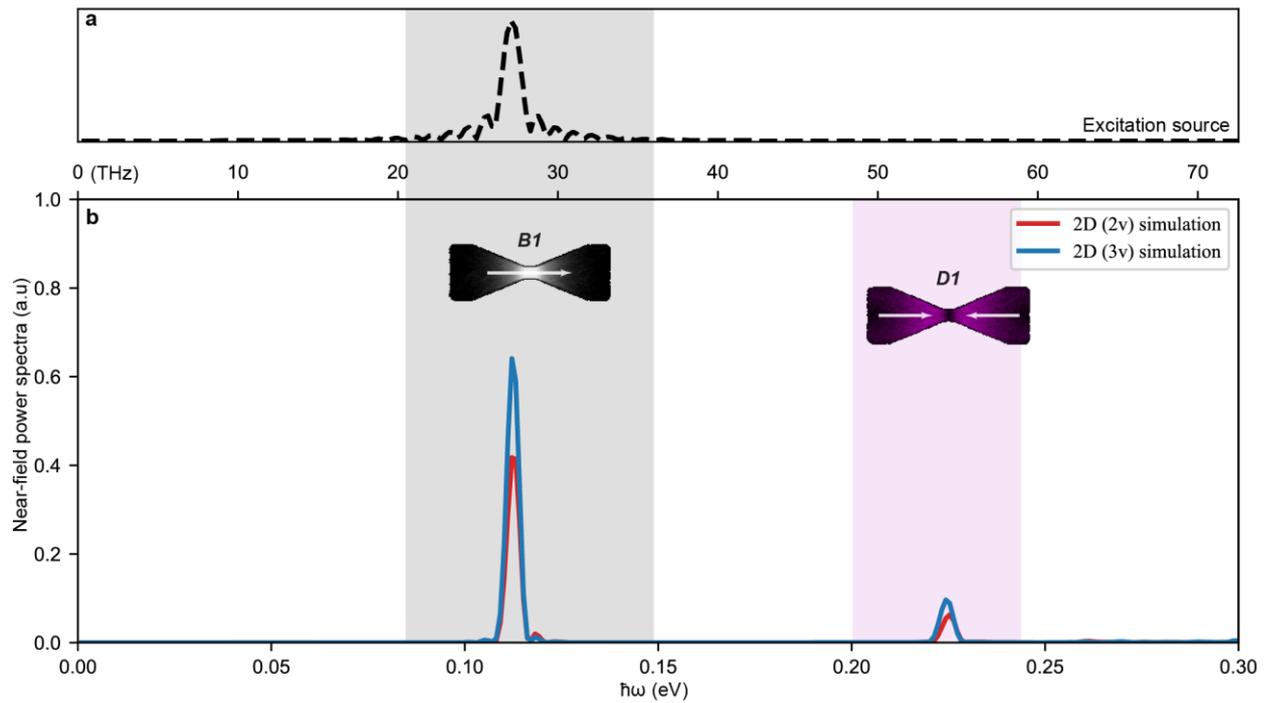

Fig. S16 Comparison between 2D (2v) and 2D (3v) simulation (parabolic-band electrons). The same structure that was discussed in Fig. 3 is excited with a narrow-band excitation at the **B1** resonance frequency. In 2D (2v) simulations, only the x and y components of the macroparticle momentum are initialized and calculated. In 2D (3v) simulations, momentum in z-direction is included.

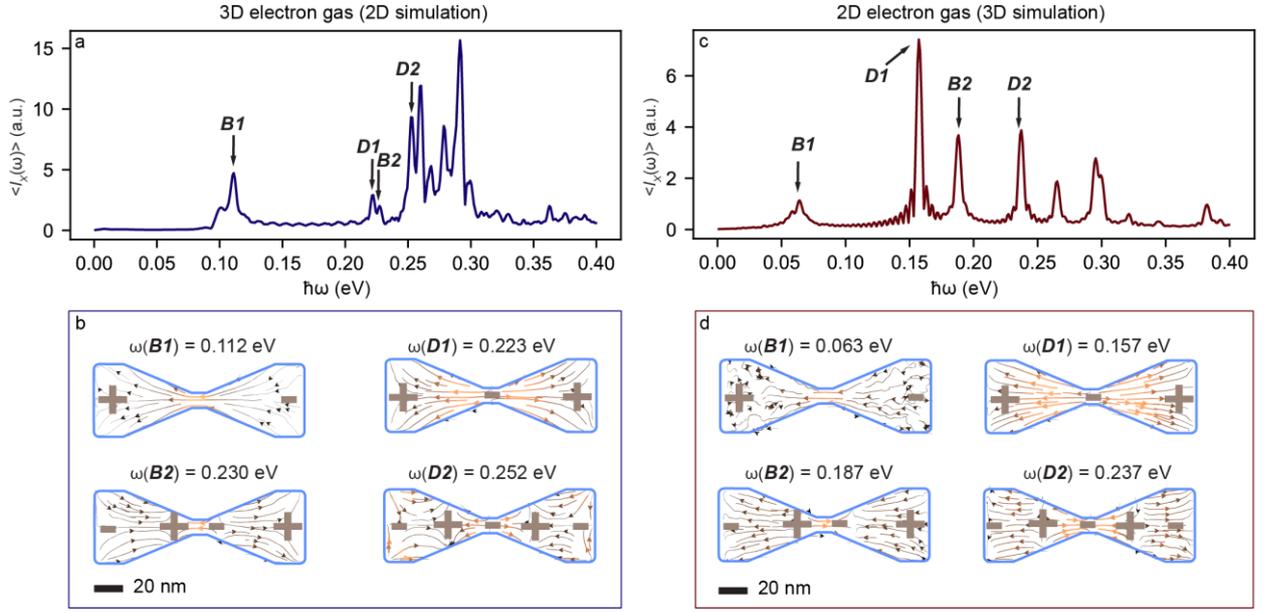

Fig. S17. Comparing the spectrum of average oscillating current $I_x(\omega)$ and mode profiles between (**a-b**) a 2D simulation of a 3D Dirac electron gas with electron density $n = 1 \times 10^{26}\ m^{-3}$ excited by a broad-band dipole and (**c-d**) a 3D simulation of a 2D Dirac electron gas in free-hanging graphene with sheet electron density $n = 10^{13}\ cm^{-2}$ excited by an electron beam moving in z-direction at v=c/2, 10 nm from the left edge of the bowtie. The bow tie geometry is kept the same as the structure simulated in Figs. 3, 4, 5 and 6. (**b**) and (**d**) show the stream plots of current density $J$ for the first 4 resonance modes of the 3D electron gas and 2D electron gas cases, respectively, with the corresponding resonance frequencies indicated on top of each map.

## 13. Inelastic mean-free-path

Inelastic scattering on impurities are simulated stochastically with the probability of an electron being scattered within a timestep $\Delta t$ is $\Delta t/\tau$ with $\tau$ being the relaxation time[4]. However, the collision only randomizes the direction of the particle while its velocity is kept constant. For most of the presented results here, the mean-free-path $\lambda \gg L$, with $L$ being the characteristic length of the resonator, which is the case for graphene. The scattering mechanism here is then omitted to save computation time. For completeness, Fig. S18 shows how second harmonic generation depends on the impurity scattering mean-free-path.

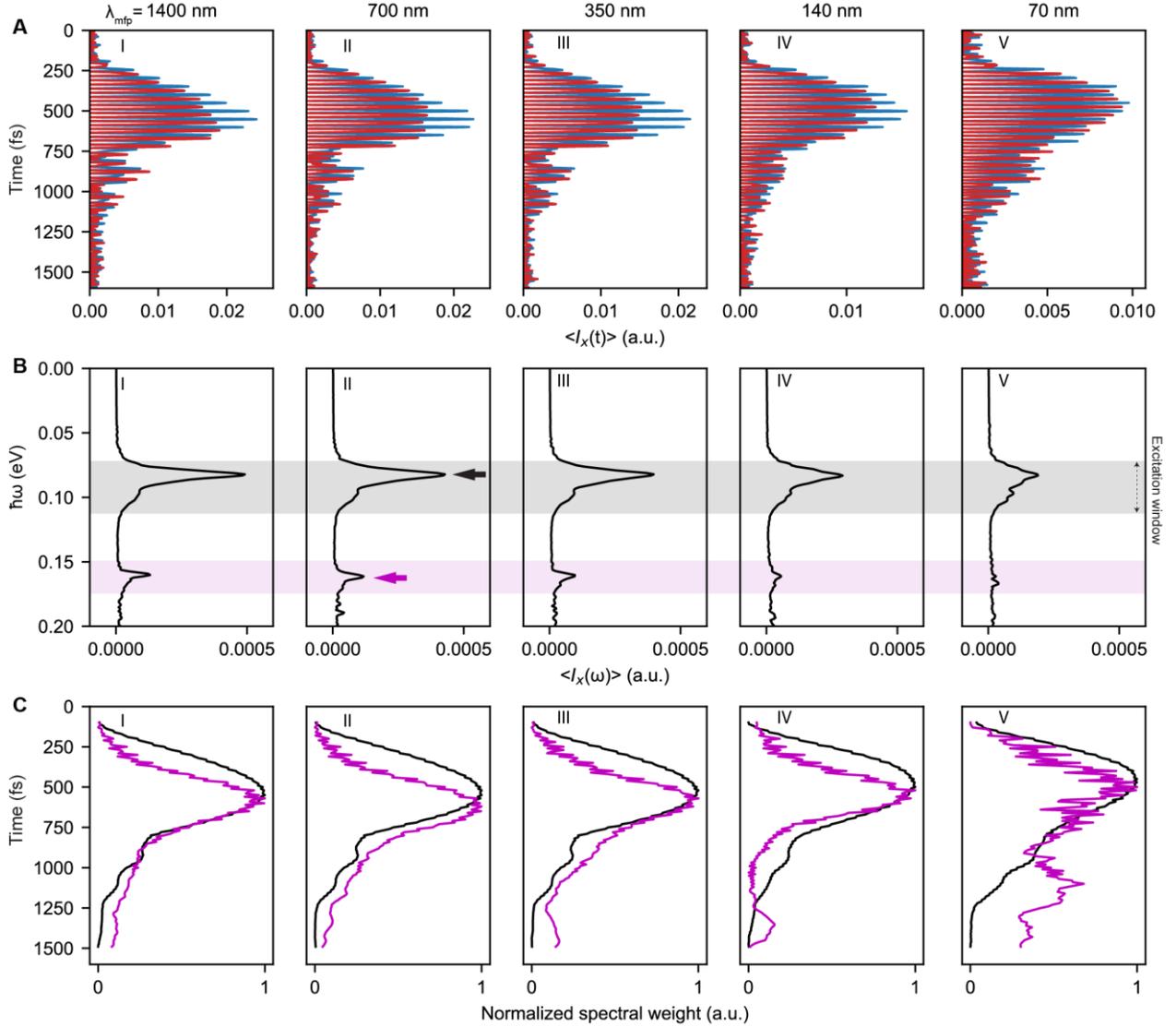

Fig. S18. Varying inelastic mean-free-path $\lambda_{mfp}$ in 160 nm long bowtie structures with a tapering angle of 23° (Dirac electrons), the same structure that was discussed in Fig. 2, 3, 4. The inelastic mean-free-path is incorporated in PIC as described in Methods. (A) Time-dependent oscillating current on the left wing of the bow tie (tapering angle of 20°) with blue and red indicating forward and backward electron flow. (B) shows the same oscillating current in frequency domain. (C) shows the time-dependent spectral weight of the fundamental peak ($\omega_1$) and the secondary harmonic peak ($\omega_2$).

## 14. More details for Fig. 6

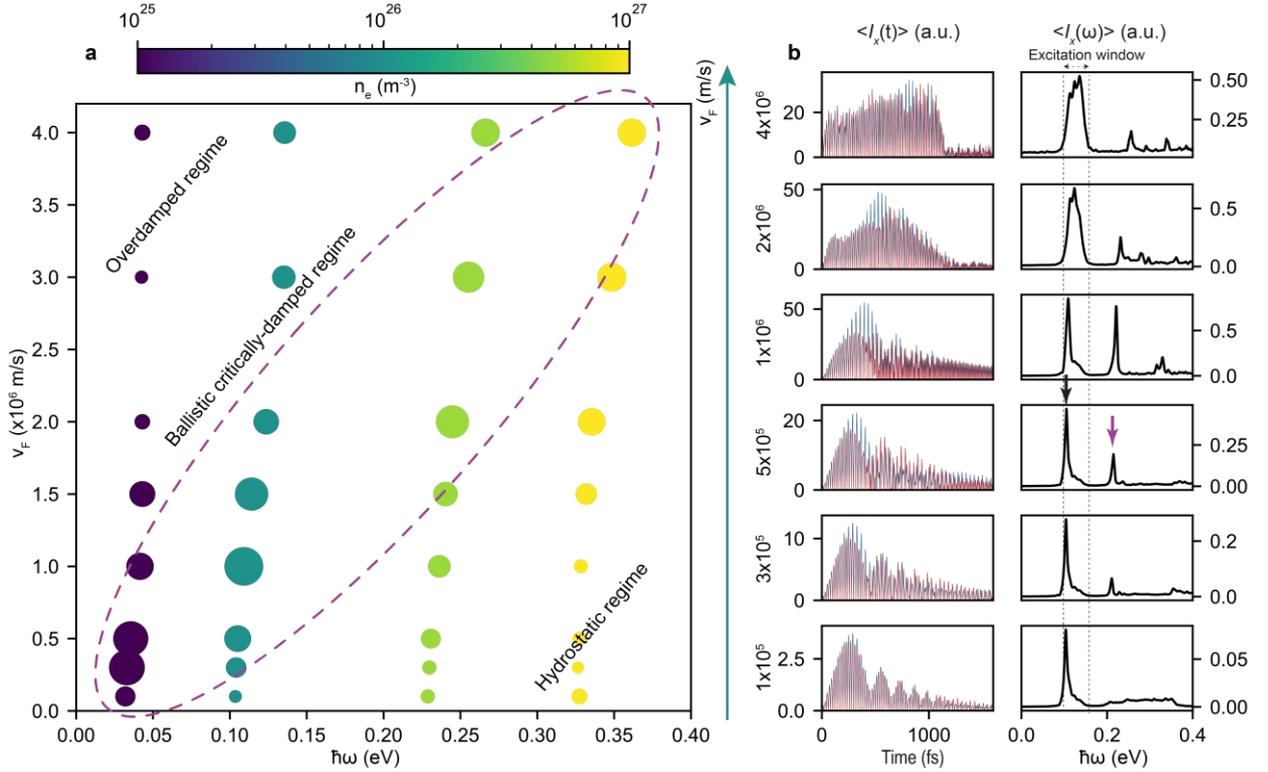

Fig. S19. Sweeping across operating regimes. (A) The same bow tie geometry as in Fig. 4 is excited with $\omega_{excitaion} = \omega_{B1} = \frac{1}{2}\omega_{D1}$. The marker size indicates the ratio between the second harmonic and the fundamental mode intensity as a function of Fermi velocity and excitation frequency. $\omega_{exc}$ is tuned by varying the electron density: for purple, $n_e$ is $10^{25}$ m$^{-3}$, for blue-green $n_e$ is $10^{26}$ m$^{-3}$, for light green $n_e$ is 5×$10^{26}$ m$^{-3}$, and for yellow $n_e$ is $10^{27}$ m$^{-3}$ (B) Response in the time domain and (C) frequency domain for systems with fixed $n_e = 1 \times 10^{26}\ m^{-3}$ while varying $v_F$. The excitation field strength is normalized to be proportional to the Fermi velocity such that the ratio $\frac{v_{\text{drift}}}{v_F}$ is consistent while varying $v_F$ for a fair comparison, accounting for the effect of field strength presented in Fig. 4.

## 15. Boundary conditions for the material surface

**Hard boundary conditions**

The material boundary is specified with a field of surface normal vectors $n_S$ defined at the nodes of the Yee-grid, with non-zero values outside the boundary of the designed structure. At each timestep after the particle pusher, the normal vector at each particle position $n_{S,\alpha}$ is interpolated with the nearest-neighbour interpolation to determine whether surface scattering is applied when $|n_{S,\alpha}| > 0$ and $n_{S,\alpha} \cdot p_{\alpha,t} > 0$.

For specular scattering, its momentum is specularly reflected:

$$p_{\alpha,t,f} = p_{\alpha,t} - 2 \frac{n_{S,\alpha} \cdot p_{\alpha,t}}{(|n_{S,\alpha}|)^2} n_{S,\alpha}$$

If diffuse scattering, we conserve the magnitude of the momentum while its direction is randomly chosen to be at an angle $\theta$ with the $n_{S,\alpha}$ with the probability density $\sim cos(\theta)$ using an inverse sampling method.

Its position is updated to keep particles within the boundary as follows:

$$x_{\alpha,t,f} = x_{\alpha,t} - \frac{n_{S,\alpha} \cdot (x_{\alpha,t} - x_{\alpha,t-\Delta t})}{(|n_{S,\alpha}|)^2} n_{S,\alpha}$$

**Soft versus Hard boundary conditions**

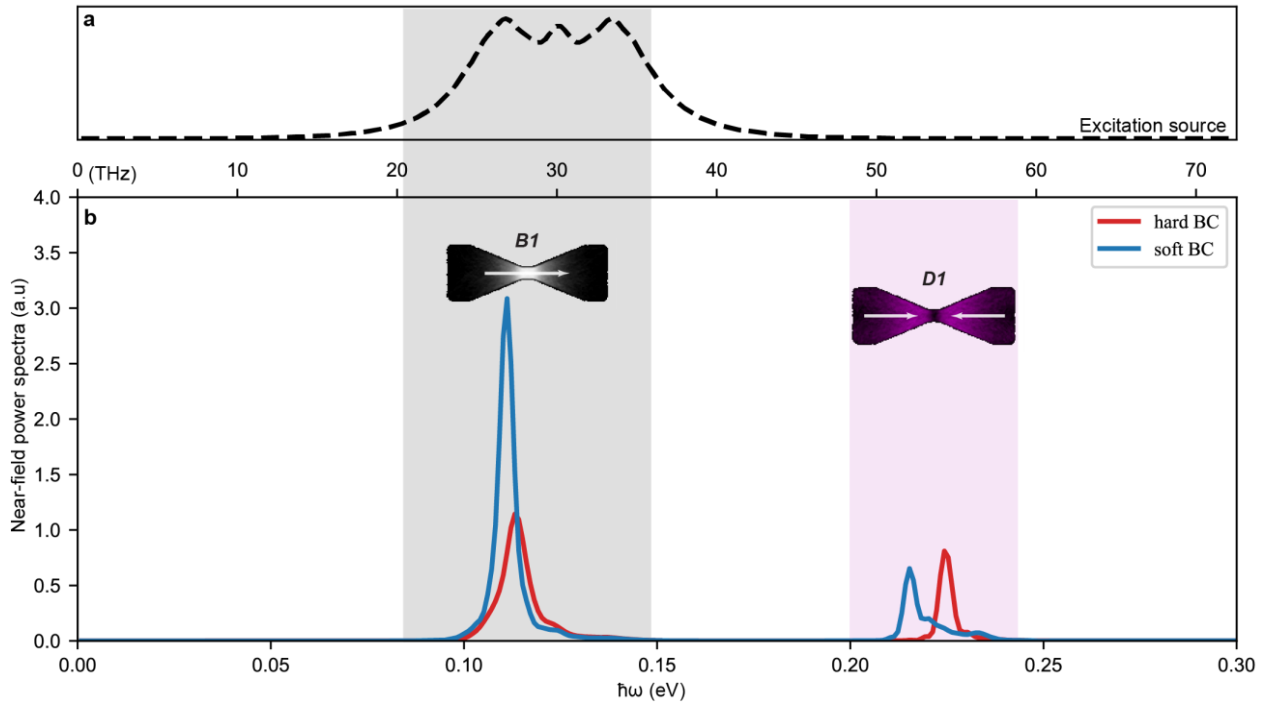

Fig. S20. Comparison between soft boundary conditions (BC) and hard boundary conditions for charge carriers. Same structure in Fig. 3, 4, 5 is excited with a broadband excitation source (**a**) In hard BC case, specular reflections at the material surface are set to occur within 1 simulation timestep as described in Methods section using a conditional BC with nearest-grid interpolation. In soft BC case, the charge carriers are reflected over several simulation timesteps by an external electric field (2V/nm) normal to the material surface. The external electric field does not participate in the Maxwell solver, interpolated at the position of charge carriers with second-order interpolation[14]. Soft boundary conditions will be relevant for geometries defined with an electrostatic barrier[15].

16. **Table S1. Simulation parameters used in the results shown in this chapter.**

The particles are initialized with a Fermi-Dirac distribution or with a fixed Fermi velocity for parabolic-band and Dirac-band charge carriers, respectively. The simulation cell-size is chosen to be 1×1 nm² in order to resolve the Debye length of the free electron plasma. Each cell is initialized with 64 macro-particles distributed regularly. The simulation box is 512×512 nm². The incoming light is polarized in the x-direction, the longitudinal axis of the bow ties. The EM boundary condition is set to be periodic for the y-direction and silver-muller[16] for the x-direction. The electron density used for the calculations in Figures 2-5 and in Figure 7 is $10^{26}$ m$^{-3}$. More details about the simulation parameters for the presented results are summarised below.

| Figure name | Carrier type | $n_e$ $(m^{-3})$ | $m_T^*, m_L^*$ | $v_F$ $(m/s)$ | Excitation frequency | Excitation field strength |
|---|---|---|---|---|---|---|
| 2b-c | Parabolic-band electrons | $10^{26}$ | $m_e, m_e$ | $10^6$ | Broadband (0.05-0.15 eV) | 20 MV/m |
| 2d | Parabolic-band electrons | $10^{26}$ | $m_e, m_e$ | $10^6$ | Narrow-band (0.112 eV) | 5 MV/m |
| 3a-d | Parabolic-band electrons/holes | $10^{26}$ | $m_e, m_e$ | $10^6$ | Narrow-band (0.112 eV) | 5 MV/m |
| 3e-h | Dirac electrons/holes | $10^{26}$ | $m_e/2, \infty$ | $10^6$ | Narrow-band (0.112 eV) | 5 MV/m |
| 4a, c | Parabolic-band electrons | $10^{26}$ | $m_e, m_e$ | $10^6$ | Narrow-band (0.112 eV) | As indicated in Fig. S12 |
| 4b, d | Dirac electrons | $10^{26}$ | $m_e/2, \infty$ | $10^6$ | Narrow-band (0.112 eV) | As indicated in Fig. S12 |
| 5 | Dirac electrons | $10^{26}$ | $m_e/2, \infty$ | $10^6$ | Narrow-band (varying frequency) | 20 MV/m |
| 6 | Dirac electrons | $10^{25} - 10^{27}$ | $m_e/2, \infty$ | $10^5 - 10^7$ | Broadband (varying range to cover B1 mode, scale with $\sqrt{n_e}$, For $n_e = 10^{26} m^{-3}$, frequency range is 0.1-0.15 eV) | 20 MV/m |
| 7a-c | Parabolic-band electrons | $10^{26}$ | $m_e, m_e$ | $10^6$ | Narrow-band (0.108 eV) | 20 MV/m |
| 7d-g | Dirac electrons | $10^{26}$ | $m_e/2, \infty$ | $10^6$ | Narrow-band (0.108 eV) | 20 MV/m |
| S1 | Parabolic-band electrons | $10^{26}$ | $m_e, m_e$ | $10^6$ | Broadband (0.1-0.15 eV) | 20 MV/m |
| S3 | Parabolic-band electrons | $10^{26}$ | $m_e, m_e$ | $10^6$ | Broadband (0.15-0.3 eV) | 20 MV/m |
| S5 | Parabolic-band electrons | $10^{26}$ | $m_e, m_e$ | $10^6$ | Broadband (0.1-0.3 eV) | 20 MV/m |
| S7, S8 | Parabolic-band electrons | $10^{26}$ | $m_e, m_e$ | $10^6$ | Narrow-band (varying frequency) | 5 MV/m |
| S15 | Parabolic-band electrons | $10^{26}$ | $m_e, m_e$ | $10^6$ | Broadband (0.05-0.15 eV) | 20 MV/m |
| S16 | Dirac electrons | $10^{26}$ | $m_e, \infty$ | $10^6$ | Broadband (0.07-0.11 eV) | 20 MV/m |
| S17a-b | Dirac electrons | $10^{26}$ | $m_e/2, \infty$ | $10^6$ | Broadband Dipole (0.1-0.3 eV) | |

| S17c-d | 3D simulation of free-hanging graphene | $10^{13}\ (cm^{-2})$ | $\frac{\hbar\sqrt{\pi n_e}}{2v_F} = 0.03m_e, \infty$ | $10^6$ | 100 e- charge moving at 0.5 c | |
|---|---|---|---|---|---|---|
| S18 | Parabolic-band electrons | $10^{26}$ | $m_e, m_e$ | $10^6$ | Narrow-band (0.112 eV) | 5 MV/m |
| S20 | Parabolic-band electrons | $10^{26}$ | $m_e, m_e$ | $10^6$ | Broadband (0.1-0.15 eV) | 20 MV/m |